\documentclass[%
 reprint,
 amsmath,amssymb,
 aps,
]{revtex4-1}

\usepackage{tikz}
\usepackage{xfrac}
\usepackage{graphicx}
\usepackage{dcolumn}
\usepackage{bm}
\usepackage{mathtools}
\usepackage[pdftex,colorlinks=true,linkcolor=blue,citecolor=blue,urlcolor=black]{hyperref}

\newcommand{\ket}[1]{| #1 \rangle}
\newcommand{\bra}[1]{\langle #1|}
\newcommand{\ip}[2]{\langle #1|#2\rangle}

\DeclareMathOperator{\poly}{poly}
\DeclareMathOperator{\tr}{tr}

\newtheorem{thm}{Theorem}

\begin{document}


\title{Quantum algorithms and the finite element method}

\author{Ashley Montanaro and Sam Pallister}
\affiliation{School of Mathematics, University of Bristol, Bristol, BS8 1TW, UK.
}%

\date{\today}

\begin{abstract}
The finite element method is used to approximately solve boundary value problems for differential equations. The method discretises the parameter space and finds an approximate solution by solving a large system of linear equations. Here we investigate the extent to which the finite element method can be accelerated using an efficient quantum algorithm for solving linear equations. We consider the representative general question of approximately computing a linear functional of the solution to a boundary value problem, and compare the quantum algorithm's theoretical performance with that of a standard classical algorithm -- the conjugate gradient method. Prior work had claimed that the quantum algorithm could be exponentially faster, but did not determine the overall classical and quantum runtimes required to achieve a predetermined solution accuracy. Taking this into account, we find that the quantum algorithm can achieve a polynomial speedup, the extent of which grows with the dimension of the partial differential equation. In addition, we give evidence that no improvement of the quantum algorithm could lead to a super-polynomial speedup when the dimension is fixed and the solution satisfies certain smoothness properties.
\end{abstract}

\maketitle



\section{Introduction}

The development of a quantum algorithm for large systems of linear equations is an exciting recent advance in the field of quantum algorithmics. First introduced by Harrow, Hassidim and Lloyd~\cite{harrow09}, and later improved by other authors~\cite{ambainis12b,childs15b}, the algorithm gives an exponential quantum speedup over classical algorithms for solving linear systems. However, the quantum linear equation (QLE) algorithm ``solves'' a system of equations $A\mathbf{x} = \mathbf{b}$ in an unusually quantum sense. The input $\mathbf{b}$ is provided as a quantum state $\ket{b}$, and the algorithm produces another state $\ket{x}$ corresponding to the desired output $\mathbf{x}$. Whether this is considered to be a reasonable definition of ``solution'' depends on the intended application~\cite{aaronson15}. Still, linear equations are so ubiquitous in science and engineering that many applications of the QLE algorithm have been proposed, ranging from machine learning~\cite{lloyd13} to computing properties of electrical networks~\cite{wang13}.

One area in which large systems of linear equations occur is the finite element method (FEM)~\cite{axelsson01,ciarlet78,brenner08,rao05}. The FEM is a technique for efficiently finding numerical approximations to the solutions of boundary value problems (BVPs) for partial differential equations, based on discretising the parameter space via a finite mesh. The FEM is a tempting target for acceleration by the QLE algorithm for several reasons. First, the large systems of linear equations that occur in the FEM are produced algorithmically, rather than being given directly as input. This avoids efficiency issues associated with needing to access data via a quantum RAM~\cite{lloyd13,aaronson15}. Second, the FEM naturally leads to sparse systems of linear equations, which is usually a requirement for quantum speedup via the QLE algorithm. Third, the FEM has many important practical applications. These include structural mechanics, thermal physics and fluid dynamics~\cite{rao05}. Any quantum speedup for the FEM would thus represent a compelling application of quantum computers.

Clader, Jacobs and Sprouse~\cite{clader13} have studied the application of the QLE algorithm to the FEM. In particular, they consider an electromagnetic scattering cross-section problem solved via the FEM, and argue that the quantum algorithm achieves an exponential speedup for this problem over the best classical algorithm known. In order to achieve this result, the authors of~\cite{clader13} propose ways to avoid issues with the QLE algorithm that can reduce or eliminate a quantum speedup. For example, they show that the important classical technique known as preconditioning, which reduces the condition number of the input matrix $A$, can be applied within the quantum algorithm.

However, the analysis of~\cite{clader13} does not fully calculate and combine all contributions to the complexity of approximately solving the scattering cross-section problem. The classical and quantum algorithmic complexity is calculated in~\cite{clader13} in terms of two parameters: $N$ (the size of the system of linear equations resulting from applying the FEM), and $\epsilon$ (the solution accuracy). The size of the system of equations is a parameter which can be chosen by the user in order to achieve a desired accuracy (i.e.\ $N$ and $\epsilon$ are formally related). In~\cite{clader13} they are treated as independent parameters and hence the complexity analysis is left incomplete. If the scaling of $N$ with $\epsilon$ is benign, the classical algorithm might not need to solve a large system of equations to achieve a given accuracy, so the quantum speedup could be reduced or even eliminated.


\subsection{New results}

\begin{table*}
\begin{tabular}{|l|c|c|}
\hline Algorithm & No preconditioning & Optimal preconditioning\\
\hline Classical & $\phantom{\bigg|}\widetilde{O}\left( (|u|_2 / \epsilon)^{(d+1)/2}\right)$ & $\widetilde{O}\left( (|u|_2/\epsilon)^{d/2}\right)$ \\
\hline Quantum & $\phantom{\bigg|}\widetilde{O} \left( \|u\| |u|_2^2/ \epsilon^3 + \|u\|_1 |u|_2/ \epsilon^2 \right)$ & $\widetilde{O} \left( \|u\|_1/ \epsilon \right)$ \\
\hline
\end{tabular}
\caption{Complexity comparison of the algorithms studied in this work. Quantities listed are the worst-case time complexities of approximating a linear functional of the solution to a $d$-dimensional BVP up to accuracy $\epsilon$, using the FEM with linear basis functions (see Section \ref{sec:quantumfem} for bounds when using higher-degree polynomials). $\|u\|$, $|u|_\ell$ and $\Vert u \Vert_\ell$ are the $L^2$ norm, Sobolev $\ell$-seminorm and Sobolev $\ell$-norm of the solution respectively, defined in Section \ref{sec:femintro}. The $\widetilde{O}$ notation hides polylogarithmic factors.}
\label{tab:results}
\end{table*}

In this paper we work through the details of applying the QLE algorithm to the general FEM, and compare the worst-case performance of the quantum algorithm with that of a simple standard classical algorithm. We choose a representative general problem -- approximating a linear functional of the solution to a BVP corresponding to an elliptic PDE -- which allows the two types of algorithm to be fairly compared.

Our results can be summarised as follows: We find that the QLE algorithm is indeed applicable to the general FEM, and can achieve substantial speedups over the classical algorithm. However, the quantum speedup obtained is only at most polynomial, if the spatial dimension is fixed and the solution satisfies certain smoothness properties. For example, the maximal advantage of the quantum algorithm for the typical physically relevant PDE defined over 3+1 dimensions (three spatial and one temporal, such that $d=4$) is approximately quadratic. In small enough dimension, and if the solution is sufficiently smooth, the runtime of the quantum algorithm can actually be worse than the classical algorithm.

Examples of the bounds we derive are listed in Table \ref{tab:results}, which includes the effect of preconditioning on the runtime of the algorithms. Note that in general it is difficult to rigorously analyse the performance of preconditioners. We therefore choose to highlight two extreme possibilities: no preconditioning at all is applied, or maximally successful preconditioning is used. The true performance of an algorithm using preconditioning will fall somewhere between the two cases. 




The runtime of both the classical and quantum algorithms depends on the Sobolev $\ell$-seminorm and Sobolev $\ell$-norm of the solution to the BVP, for some $\ell$; roughly speaking, these measure the size of the $\ell$'th derivatives of the solution. Assuming that preconditioning has been optimally used within the QLE algorithm, the quantum algorithm's runtime is dependent only on the Sobolev 1-norm (up to polylogarithmic terms). However, the classical algorithm's runtime depends on the Sobolev $\ell$-seminorm, for some $\ell \ge 2$. Therefore, for problems with solutions whose higher-order derivatives are large, the quantum advantage could be substantial.

Perhaps more importantly, to achieve accuracy $\epsilon$ in spatial dimension $d$, the runtime of the classical algorithm scales as $\epsilon^{-O(d)}$, while the scaling with $\epsilon$ of the quantum algorithm's runtime does not depend on $d$. For higher-dimensional problems, the quantum speedup can thus be very significant.  Interestingly, this holds even if preconditioning is not used. (Note that we cannot quite say that the quantum algorithm achieves an {\em exponential} speedup, as the runtime also contains a dimension-dependent constant factor which could be very large.)

One example application is any dynamical problem involving $n$ bodies, which implies solving a PDE defined over a configuration space of dimension $2n$.
Also, there may be a significant advantage for problems in mathematical finance; for example, pricing multi-asset options requires solving the Black-Scholes equation over a domain with dimension given by the number of assets~\cite{jiang05}.
This is discussed further in the conclusion to this paper.

The reason for the apparent contradiction between our results and previous work~\cite{clader13}, which claimed an exponential speedup in fixed spatial dimension, is the inclusion of an accuracy parameter in the runtime, which was not fully incorporated in~\cite{clader13}. Imagine that we would like to produce a solution to some BVP that is accurate up to $\epsilon$. This accuracy parameter will affect the runtime of algorithms for the FEM. There are two potential sources of error in producing the solution: the discretisation process which converts the problem to a system of linear equations, and any inaccuracies in solving the system of equations itself and computing the desired function of the solution. The larger the system of equations produced, the smaller the first type of error is.

The QLE algorithm can work with an exponentially larger set of equations in a comparable time to the classical algorithm, so this source of error can be reduced exponentially. However, the scaling with accuracy of the QLE algorithm's extraction of a solution from the system of linear equations is substantially {\em worse} than the classical algorithm. These two effects can come close to cancelling each other out.

We remark that there is a subtle point here: the scaling with accuracy of the quantum algorithm is substantially better if we only wish to produce the quantum state corresponding to the solution to the FEM~\cite{childs15b}, rather than computing some property of the state by measuring it. However, in applications one will always eventually want to perform a measurement to extract information from the final output of the quantum algorithm. We therefore consider it reasonable to compare the quantum and classical complexities of producing a (classical) answer to some given problem.

Finally, we argue that the inability of the quantum algorithm to deliver exponential speedups (in some cases) is not a limitation of the algorithm itself, but rather that any quantum algorithm for the FEM will face similar constraints. We elucidate several barriers with which any quantum algorithm will have to contend. First, we show that, informally, any algorithm which needs to distinguish between two states which are distance $\epsilon$ apart must have runtime $\Omega(1/\sqrt{\epsilon})$. Second, we argue that the ``FEM solving subroutine'' of any quantum algorithm can likely be replaced with an equivalent classical subroutine with at most a polynomial slowdown (in fixed spatial dimension and when the solution is smooth). Third, we show that there can be no more than a quadratic speedup if the input to the problem is arbitrary and accessed via queries to a black box or ``oracle''.

Our results pinpoint the regimes in which one can hope to achieve exponential quantum speedups for the FEM, and show that apparent speedups can disappear when one takes the effect of solution accuracy into account. Nevertheless, we believe that the fact that exponential speedups might still be obtained in some cases is encouraging, and an incentive to focus on problems with a possibility of a genuine exponential quantum speedup.


\subsection{Other related work}

An alternative approach to the approximate numerical solution of PDEs is the finite difference method (FDM). This method is also based on discretisation of the problem domain, but differs from the FEM in that it approximates the partial derivatives in the original problem with finite differences.

The QLE algorithm can also be applied to the FDM. One example where this has been done, and described in detail, is work of Cao et al.~\cite{cao13}, who gave a quantum algorithm for the Poisson equation in $d$ dimensions. Their algorithm produces a quantum state corresponding to the solution to the equation in time $O(\max\{d,\log 1/\epsilon\} \log^3 1/\epsilon)$. Note that this scaling with $\epsilon$ is exponentially better than the best general results on Hamiltonian simulation known at the time; their algorithm used special properties of the Poisson equation to achieve an improved runtime. The best classical algorithms require time $\epsilon^{-\Omega(d)}$ as they solve a discretised version of the problem on a $d$-dimensional grid with cells of size $\epsilon \times \epsilon \times \dots \times \epsilon$.

However, the quantum algorithm of Cao et al.~\cite{cao13} shares the property of the FEM algorithms discussed here that, in order to extract some information from the quantum state produced, one finishes with a scaling with $\epsilon$ which is $\poly(1/\epsilon)$. In the physically realistic setting of the dimension $d$ being fixed and the accuracy $\epsilon$ being the parameter of interest, this is only a polynomial improvement.

Other related work has given quantum algorithms for solving large systems of sparse linear~\cite{berry14} or nonlinear~\cite{leyton08} differential equations via Euler's method. In these cases the quantum algorithms can in principle achieve an exponential improvement over classical computation for approximately computing properties of the solution to the system, if the system of equations is provided implicitly. Fleshing out this approach requires also specifying how the equations are produced and how the property of interest is computed. If the equations are generated by a discretisation procedure such as the FDM, similar qualitative conclusions to those we derive for the FEM seem likely to hold.


\subsection{Organisation and notation}

We begin, in Section~\ref{sec:femintro}, by introducing the FEM and describing its classical complexity. Section~\ref{sec:quantumfem} goes through the details of applying the QLE algorithm to the FEM and determines its complexity. In Section~\ref{sec:quantumlower} we describe various limitations on the quantum algorithm. We conclude in Section~\ref{sec:conclusions} with some discussion and open problems.

We will need to deal with continuous functions, their discretised approximations as vectors, and their corresponding quantum states. Italics denote functions, boldface denotes vectors, and quantum states (usually normalised) are represented as kets. We often let $\Omega \subseteq \mathbb{R}^d$ denote an arbitrary convex set. For a function $f \in L^2(\Omega)$, $\|f\| := \left(\int_\Omega f(x)^2 dx \right)^{1/2}$ denotes the $L^2$ norm of $f$. For a vector $\mathbf{f}$, $\|\mathbf{f}\| := \left(\sum_i \mathbf{f}_i^2\right)^{1/2}$ denotes the $\ell_2$ norm of $\mathbf{f}$.

We often use the term ``spatial dimension'' as shorthand for ``number of degrees of freedom in the given PDE'', and as distinct from the dimension of the vector space used for a discretised approximation of the solution of a PDE, or the dimension of the Hilbert space acted on by a quantum algorithm; this is merely for convenience and should not be taken to mean that the only PDEs of interest are those in which the degrees of freedom are physical spatial dimensions.


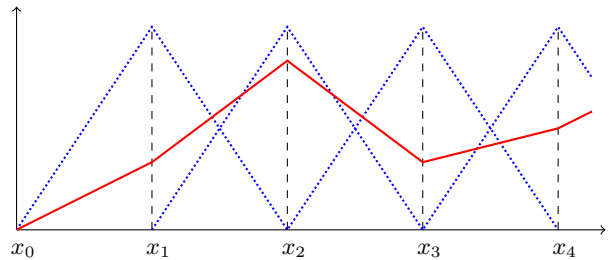
\begin{figure}
\centering
\begin{tikzpicture}[scale=0.9]
\draw[->] (0,0) -- (8.7,0);
\draw[->] (0,0) -- (0,3.3);
\draw[dashed] (2,0) -- (2,3);
\draw[dashed] (4,0) -- (4,3);
\draw[dashed] (6,0) -- (6,3);
\draw[dashed] (8,0) -- (8,3);
\draw[thick, blue, densely dotted] (0,0) -- (2,3) -- (4,0) -- (6,3) -- (8,0);
\draw[thick, blue, densely dotted] (2,0) -- (4,3) -- (6,0) -- (8,3) -- (8.5,2.25);
\draw[thick, red] (0,0) -- (2,1) -- (4,2.5) -- (6,1) -- (8,1.5) -- (8.5,1.75);
\node at (0.1,-0.3) {$x_0$};
\node at (2.1,-0.3) {$x_1$};
\node at (4.1,-0.3) {$x_2$};
\node at (6.1,-0.3) {$x_3$};
\node at (8.1,-0.3) {$x_4$};
\end{tikzpicture}
\caption{A basis set of ``tent'' functions (the blue, or dotted, lines), for piecewise linear functions defined on the line (an example of which is given by the red, or solid, line). Any piecewise linear function can be uniquely specified as a sum of scaled tents.}
\label{fig:tent}
\end{figure}

\section{The finite element method}
\label{sec:femintro}

Rather than provide a formal introduction to the FEM, it is easiest to motivate the procedure via an example (we refer the reader to \cite{ciarlet78,brenner08,axelsson01} for a thorough treatment). Imagine we would like to solve Poisson's equation on the interval $[0,1]$, in one dimension:
\[ u'' = f; \quad u(0)=u'(1)=0. \]
Here $f$ is the input to the problem and we fix the boundary conditions $u(0)$, $u'(1)$. Given a sufficiently smooth ``test function'' $v \in L^2[0,1]$ such that $v(0)=0$, one can multiply both sides by $v$ and then integrate by parts:
\[ \int_0^1 f(x)v(x) dx = \int_0^1 u''(x)v(x) dx = - \int_0^1 u'(x)v'(x) dx. \]
Assuming certain regularity properties of $f$, a function $u$ which satisfies this equality for all test functions $v$ will satisfy Poisson's equation. This is known as the \emph{weak formulation} of Poisson's equation. The goal is to reduce this formulation to a problem that is tractable computationally. The approximation is to consider solutions and test functions that instead exist in some finite-dimensional subspace $S$ of $L^2[0,1]$. Denote the approximate solution as $\tilde{u}$, such that $\tilde{u} \in S \subset L^2[0,1]$. Commonly $S$ is taken to be the space of piecewise polynomial functions of some degree $k$; the choice of ``pieces'' for these functions is the origin of the finite element mesh.

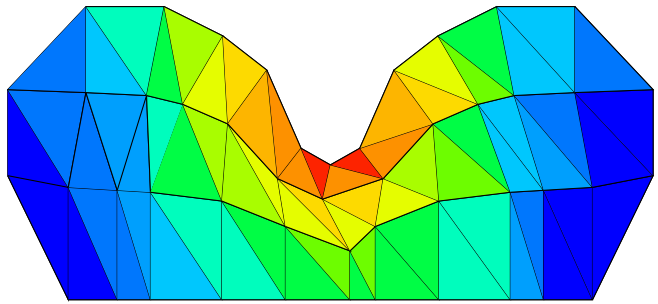
\begin{figure}
\centering
\begin{tikzpicture}[yscale=-1, scale=0.13]
    \definecolor{dark_red}{RGB}{253,35,0}
    \definecolor{orange}{RGB}{253,145,0}
    \definecolor{light_orange}{RGB}{253,181,0}
    \definecolor{yellow_orange}{RGB}{253,218,0}
    \definecolor{yellow}{RGB}{229,253,0}
    \definecolor{yellow_green}{RGB}{169,253,0}
    \definecolor{light_green}{RGB}{109,253,0}
    \definecolor{turquoise}{RGB}{0,253,69}
    \definecolor{cyan}{RGB}{0,253,189}
    \definecolor{light_blue}{RGB}{0,199,253}
    \definecolor{royal_blue}{RGB}{0,159,253}
    \definecolor{dark_blue}{RGB}{0,119,253}
    
    \draw (0.0,8.5) -- (8.0,8.8) -- (14.2,9.1) -- (17.9,10.0) -- (22.5,12.0) -- (27.6,17.6) -- (32.2,19.7) -- (38.4,17.6) -- (43.5,12.0) -- (48.1,10.0) -- (51.8, 9.1) -- (58.0, 8.8) -- (66.0,8.5);
    \draw (0.0,17.3) -- (6.2,18.5) -- (11.2,18.8) -- (14.6,19.0) -- (21.9,19.9) -- (28.4,22.5) -- (35.0,25.0) -- (37.6,22.5) -- (44.1,19.9) -- (51.4,19.0) -- (54.8,18.8) -- (59.8,18.5) -- (66.0,17.3);
    \draw (0.0,17.3) -- (6.2,30.0) -- (11.2,30.0) -- (14.6,30.0) -- (21.9,30.0) -- (28.4,30.0) -- (35.0,30.0) -- (37.6,30.0) -- (44.1,30.0) -- (51.4,30.0) -- (54.8,30.0) -- (59.8,30.0) -- (66.0,17.3);
    \draw ( 8.0, 0.0) -- ( 0.0, 8.5) -- ( 0.0,17.3);
    \draw ( 8.0, 0.0) -- ( 8.0, 8.8) -- ( 6.2,18.5) -- ( 6.2,30.0);
    \draw (16.0, 0.0) -- (14.2, 9.1) -- (11.2,18.8) -- (11.2,30.0);
    \draw (22.0, 3.0) -- (17.9,10.0) -- (14.6,19.0) -- (14.6,30.0);
    \draw (26.5, 6.5) -- (22.5,12.0) -- (21.9,19.9) -- (21.9,30.0);
    \draw (30.0,14.5) -- (27.6,17.6) -- (28.4,22.5) -- (28.4,30.0);
    \draw (33.0,16.2) -- (32.2,19.7) -- (35.0,25.0) -- (35.0,30.0);
    \draw (36.0,14.5) -- (38.4,17.6) -- (37.6,22.5) -- (37.6,30.0);
    \draw (39.5, 6.5) -- (43.5,12.0) -- (44.1,19.9) -- (44.1,30.0);
    \draw (44.0, 3.0) -- (48.1,10.0) -- (51.4,19.0) -- (51.4,30.0);
    \draw (50.0, 0.0) -- (51.8, 9.1) -- (54.8,18.8) -- (54.8,30.0);
    \draw (58.0, 0.0) -- (58.0, 8.8) -- (59.8,18.5) -- (59.8,30.0);
    \draw (66.0, 8.5) -- (66.0,17.3);
    \draw ( 8.0, 0.0) -- (16.0, 0.0) -- (22.0, 3.0) -- (26.5,6.5) -- (30.0,14.5) -- (33.0,16.2) -- (36.0,14.5) -- (39.5,6.5) -- (44.0,3.0) -- (50.0,0.0) -- (58.0,0.0) -- (66.0, 8.5);
    
    \filldraw[fill=blue, draw=black] (0.0,17.3) -- (6.2,30.0) -- (6.2,18.5);
    \filldraw[fill=blue, draw=black] (6.2,30.0) -- (11.2,30.0) -- (6.2,18.5);
    \filldraw[fill=dark_blue, draw=black] (11.2,30.0) -- (11.2,18.8) -- (6.2,18.5);
    \filldraw[fill=dark_blue, draw=black] (11.2,30.0) -- (14.6,30.0) -- (11.2,18.8);
    \filldraw[fill=royal_blue, draw=black] (14.6,30.0) -- (14.6,19.0) -- (11.2,18.8);
    \filldraw[fill=light_blue, draw=black] (14.6,30.0) -- (21.9,30.0) -- (14.6,19.0);
    \filldraw[fill=cyan, draw=black] (21.9,30.0) -- (21.9,19.9) -- (14.6,19.0);
    \filldraw[fill=cyan, draw=black] (21.9,30.0) -- (28.4,30.0) -- (21.9,19.9);
    \filldraw[fill=turquoise, draw=black] (28.4,30.0) -- (28.4,22.5) -- (21.9,19.9);
    \filldraw[fill=turquoise, draw=black] (28.4,30.0) -- (35.0,30.0) -- (28.4,22.5);
    \filldraw[fill=light_green, draw=black] (35.0,30.0) -- (35.0,25.0) -- (28.4,22.5);
    \filldraw[fill=turquoise, draw=black] (35.0,30.0) -- (37.6,30.0) -- (35.0,25.0);
    \filldraw[fill=light_green, draw=black] (37.6,30.0) -- (37.6,22.5) -- (35.0,25.0);
    \filldraw[fill=turquoise, draw=black] (37.6,30.0) -- (44.1,30.0) -- (37.6,22.5);
    \filldraw[fill=turquoise, draw=black] (44.1,30.0) -- (44.1,19.9) -- (37.6,22.5);
    \filldraw[fill=cyan, draw=black] (44.1,30.0) -- (51.4,30.0) -- (44.1,19.9);
    \filldraw[fill=cyan, draw=black] (51.4,30.0) -- (51.4,19.0) -- (44.1,19.9);
    \filldraw[fill=royal_blue, draw=black] (51.4,30.0) -- (54.8,30.0) -- (51.4,19.0);
    \filldraw[fill=dark_blue, draw=black] (54.8,30.0) -- (54.8,18.8) -- (51.4,19.0);
    \filldraw[fill=blue, draw=black] (54.8,30.0) -- (59.8,30.0) -- (54.8,18.8);
    \filldraw[fill=blue, draw=black] (59.8,30.0) -- (59.8,18.5) -- (54.8,18.8);
    \filldraw[fill=blue, draw=black] (59.8,30.0) -- (66.0,17.3) -- (59.8,18.5);
    
    \filldraw[fill=blue, draw=black] (0.0,17.3) -- (6.2,18.5) -- (0.0,8.5);
    \filldraw[fill=dark_blue, draw=black] (6.2,18.5) -- (8.0,8.8) -- (0.0,8.5);
    \filldraw[fill=dark_blue, draw=black] (6.2,18.5) -- (8.0,8.8) -- (11.2,18.8);
    \filldraw[fill=royal_blue, draw=black] (11.2,18.8) -- (8.0,8.8) -- (14.2,9.1);
    \filldraw[fill=royal_blue, draw=black] (11.2,18.8) -- (14.2,9.1) -- (14.6,19.0);
    \filldraw[fill=cyan, draw=black] (14.6,19.0) -- (14.2,9.1) -- (17.9,10.0);
    \filldraw[fill=turquoise, draw=black] (14.6,19.0) -- (21.9,19.9) -- (17.9,10.0);
    \filldraw[fill=yellow_green, draw=black] (21.9,19.9) -- (22.5,12.0) -- (17.9,10.0);
    \filldraw[fill=yellow_green, draw=black] (21.9,19.9) -- (28.4,22.5) -- (22.5,12.0);
    \filldraw[fill=yellow, draw=black] (28.4,22.5) -- (27.6,17.6) -- (22.5,12.0);
    \filldraw[fill=yellow, draw=black] (28.4,22.5) -- (35.0,25.0) -- (27.6,17.6);
    \filldraw[fill=yellow_orange, draw=black] (35.0,25.0) -- (32.2,19.7) -- (27.6,17.6);
    \filldraw[fill=yellow, draw=black] (35.0,25.0) -- (37.6,22.5) -- (32.2,19.7);
    \filldraw[fill=yellow_orange, draw=black] (37.6,22.5) -- (38.4,17.6) -- (32.2,19.7);
    \filldraw[fill=yellow, draw=black] (37.6,22.5) -- (44.1,19.9) -- (38.4,17.6);
    \filldraw[fill=yellow_green, draw=black] (44.1,19.9) -- (43.5,12.0) -- (38.4,17.6);
    \filldraw[fill=light_green, draw=black] (44.1,19.9) -- (51.4,19.0) -- (43.5,12.0);
    \filldraw[fill=turquoise, draw=black] (51.4,19.0) -- (48.1,10.0) -- (43.5,12.0);
    \filldraw[fill=light_blue, draw=black] (51.4,19.0) -- (54.8,18.8) -- (48.1,10.0);
    \filldraw[fill=light_blue, draw=black] (54.8,18.8) -- (51.8,9.1) -- (48.1,10.0);
    \filldraw[fill=royal_blue, draw=black] (54.8,18.8) -- (59.8,18.5) -- (51.8,9.1);
    \filldraw[fill=dark_blue, draw=black] (59.8,18.5) -- (58.0,8.8) -- (51.8,9.1);
    \filldraw[fill=blue, draw=black] (59.8,18.5) -- (66.0,17.3) -- (58.0,8.8);
    \filldraw[fill=blue,draw=black] (66.0,17.3) -- (66.0,8.5) -- (58.0,8.8);
    
    \filldraw[fill=dark_blue, draw=black] (0.0,8.5) -- (8.0,8.8) -- (8.0,0.0);
    \filldraw[fill=light_blue, draw=black] (8.0,8.8) -- (14.2,9.1) -- (8.0,0.0);
    \filldraw[fill=cyan, draw=black] (14.2,9.1) -- (16.0,0.0) -- (8.0,0.0);
    \filldraw[fill=turquoise, draw=black] (14.2,9.1) -- (17.9,10.0) -- (16.0,0.0);
    \filldraw[fill=yellow_green, draw=black] (17.9,10.0) -- (22.0,3.0) -- (16.0,0.0);
    \filldraw[fill=yellow,draw=black] (17.9,10.0) -- (22.5,12.0) -- (22.0,3.0);
    \filldraw[fill=yellow_orange, draw=black] (22.5,12.0) -- (26.5,6.5) -- (22.0,3.0);
    \filldraw[fill=light_orange, draw=black] (22.5,12.0) -- (27.6,17.6) -- (26.5,6.5);
    \filldraw[fill=orange, draw=black] (27.6,17.6) -- (30.0,14.5) -- (26.5,6.5);
    \filldraw[fill=orange, draw=black] (27.6,17.6) -- (32.2,19.7) -- (30.0,14.5);
    \filldraw[fill=dark_red, draw=black] (32.2,19.7) -- (33.0,16.2) -- (30.0,14.5);
    \filldraw[fill=orange, draw=black] (32.2,19.7) -- (38.4,17.6) -- (33.0,16.2);
    \filldraw[fill=dark_red, draw=black] (38.4,17.6) -- (36.0,14.5) -- (33.0,16.2);
    \filldraw[fill=orange, draw=black] (38.4,17.6) -- (43.5,12.0) -- (36.0,14.5);
    \filldraw[fill=light_orange, draw=black] (43.5,12.0) -- (39.5,6.5) -- (36.0,14.5);
    \filldraw[fill=yellow_orange, draw=black] (43.5,12.0) -- (48.1,10.0) -- (39.5,6.5);
    \filldraw[fill=yellow, draw=black] (48.1,10.0) -- (44.0,3.0) -- (39.5,6.5);
    \filldraw[fill=light_green, draw=black] (48.1,10.0) -- (51.8,9.1) -- (44.0,3.0);
    \filldraw[fill=turquoise, draw=black] (51.8,9.1) -- (50.0,0.0) -- (44.0,3.0);
    \filldraw[fill=light_blue, draw=black] (51.8,9.1) -- (58.0,8.8) -- (50.0,0.0);
    \filldraw[fill=light_blue, draw=black] (58.0,8.8) -- (58.0,0.0) -- (50.0,0.0);
    \filldraw[fill=dark_blue, draw=black] (58.0,8.8) -- (66.0,8.5) -- (58.0,0.0);
\end{tikzpicture}
\caption{An example of a ``mesh'' -- a discretisation of the domain over which the PDE is defined. Each polygon is a ``finite element'', with basis functions defined upon them. The shading of each polygon here represents the amplitude associated with the function supported on each finite element. As a physical example, the diagram could represent the material stress on a plate, induced by a deformation by a rod.}
\label{fig:mesh}
\end{figure}

A particularly simple choice of basis for this example is the space of piecewise linear functions on $[0,1]$, divided up into $N$ intervals of size $h$. A basis for this space is the set of ``tent'' functions (see Fig.~\ref{fig:tent}), defined as
\[
 \phi_i(x) = \begin{dcases*}
        \frac{1}{h}(x-x_{i-1})  & if $x \in [x_{i-1},x_i]$\\
        \frac{1}{h}(x_{i+1}-x) & if $x \in [x_i, x_{i+1}]$\\
        0 & otherwise.
        \end{dcases*}
\]
More generally, consider some choice of basis for the space $S$, denoted by $B = \{\phi_i\}$, such that $|B| = N$. We choose a basis such that $\phi_i(0) = \phi_i'(1) = 0$, so that every function in $S$ satisfies the boundary conditions. Then $\tilde{u}$ can be expanded in this basis: $\tilde{u} = \sum_j U_j \phi_j$. The corresponding weak formulation of Poisson's equation is
\[ -\sum_j U_j \int_0^1 \phi'_j(x) v'(x) dx = \int_0^1 f(x) v(x) \,dx.\]
For this condition to hold for all $v \in S$, it is sufficient for it to hold on all basis functions $\phi_i$:
\[ -\sum_j U_j \int_0^1 \phi_j'(x) \phi_i'(x) dx = \int_0^1 f(x) \phi_i(x) \,dx.\]
If we define $N$-dimensional vectors $\tilde{\mathbf{u}}$ and $\tilde{\mathbf{f}}$ such that
\[ \tilde{\mathbf{u}}_i = U_i,\;\;\;\; \tilde{\mathbf{f}}_i = \int_0^1 f(x) \phi_i(x) \, dx\]
and an $N \times N$ matrix $M$ such that
\begin{equation} \label{eq:m} M_{ij} = \int_0^1 \phi'_i(x) \phi'_j(x) dx ,\end{equation}
then the approximate solution to Poisson's equation can be determined by solving the linear system
\begin{equation} \label{eq:muf} M\tilde{\mathbf{u}} = \tilde{\mathbf{f}}. \end{equation}
This general procedure (expressing the PDE in the weak formulation, choosing a finite element mesh and basis functions and solving the resultant linear system of equations) can be extended to far more complicated PDEs, domains and boundary conditions. In higher spatial dimensions, the above framework can be naturally generalised as follows. The uniform division of $[0,1]$ into intervals is replaced with a suitably regular division of the domain into a mesh, whose elements are usually polygons (for example, triangles) or polyhedra. An example of a mesh is shown in Fig.~\ref{fig:mesh}. The space $S$ is replaced with the space of piecewise polynomials of degree $k$ on the elements of the mesh, with a basis $\{ \phi_i \}$ of polynomials supported only on adjacent mesh elements. Finally, the matrix $M$ defined in (\ref{eq:m}) is modified such that $M_{ij} = a(\phi_i,\phi_j)$, where $a(u,v)$ is an inner product depending on the PDE in question.

Here we choose not to specify which PDE we wish to solve, as the details of this procedure for particular PDEs will not be very significant when making a general comparison of quantum and classical algorithms for the FEM. However, we will restrict to elliptic second-order PDEs throughout, to avoid some technical complications. Even with this restriction, the following analysis captures many examples of physical interest; for example, electrostatics, subsonic fluid dynamics and linear elasticity.

\subsection{Comparing quantum and classical algorithms for the FEM}

The goal of this paper is to compare the performance of quantum and classical algorithms for solving BVPs via the FEM. However, the quantum algorithm does not allow the full solution $u$ to a given BVP to be obtained, but does allow certain properties of $u$ to be approximately computed. In order to fairly compare classical and quantum algorithms for solving general BVPs, we consider the representative problem of computing a linear functional of $u$. That is, for some known function $r: \Omega \rightarrow \mathbb{R}$, where $\Omega \subset \mathbb{R}^d$, we seek to compute
\[ \langle r, u \rangle := \int_\Omega r(\mathbf{x}) u(\mathbf{x}) d\mathbf{x}. \]
This is one of the simplest properties of $u$ one could hope to access. In general, we do not have complete knowledge of $u$, but have some approximation $\tilde{u}$. Although there are many sensible norms with which one could measure the quality of this approximation, one natural choice is the $L^2$ norm $\|f\| := \left( \int_\Omega f(\mathbf{x})^2 d\mathbf{x})^2\right)^{1/2}$. Then
\[ |\langle r, \tilde{u} \rangle - \langle r, u \rangle| = |\langle r, (\tilde{u} - u) \rangle| \le \|r\| \|\tilde{u} - u\| \]
by Cauchy-Schwarz. Hence an accuracy of $\epsilon$ in $L^2$ norm in an approximation of $u$ translates into an additive error of at most $\epsilon\|r\|$ in an approximation of $\langle r, u \rangle$. Therefore, approximating $\langle r, u \rangle$ up to accuracy $\epsilon \|r\|$ will be the prototypical problem considered throughout.


\subsection{Approximation errors}

If $u$ is the exact solution to a BVP, henceforth let $\tilde{u}$ be the continuous, exact solution corresponding to the discretised problem (\ref{eq:muf}); $\tilde{u}$ is the solution that a perfect linear-system solver would find. In general, however, the linear-system solver is iterative and so will not truly reach $\tilde{u}$; so also let $\widetilde{\tilde{u}}$ be the continuous, approximate solution generated by the linear-system solver.

Crucially, one can show that $\tilde{u}$ can be made quite close to $u$ by taking a sufficiently fine mesh. Indeed, consider a second order differential equation defined over a polygonal, $d$-dimensional domain (or equivalently, define $d$ as the number of degrees of freedom in the PDE). Then, take an infinite, ordered family of progressively finer meshes $\{\mathcal{M}_r\}_{r=1}^{\infty}$, constructed from a triangulation of the domain with simplices of dimension $d$. Let $k$ be the total degree of the polynomials used as basis functions. We assume throughout that both $d$ and $k$ are fixed. Given a parameter $m \in \{0,1\}$, and provided that $d > 2(k-m)$, that all angles in the mesh are bounded below by some fixed value, and that the greatest edge length $h$ in the mesh goes to zero, then the following bound is known (\cite{ciarlet78}, Thm.\ 3.2.1):
\begin{equation}\label{eq:disc_error}
|u - \tilde{u}|_m \leq C h^{k+1-m} |u|_{k+1}, \quad h \rightarrow 0,
\end{equation}
assuming that weak derivatives of $u$ of order $m$ exist. Here $C$ is a constant, independent of $h$ (but not necessarily independent of $d$ or the definition of the mesh).  $|\cdot|_m$ is the Sobolev seminorm
\[ |v|_m := \left(\sum_{\alpha,|\alpha| = m} \|\partial^\alpha v\|^2 \right)^{1/2}. \]
Here $\alpha = (\alpha_1,\dots,\alpha_d)$ is a multi-index, $|\alpha| := \sum_i \alpha_i$, and $\partial^\alpha := \left(\frac{\partial}{\partial x_1}\right)^{\alpha_1} \dots \left(\frac{\partial}{\partial x_d}\right)^{\alpha_d}$. That is, the sum is over all partial derivatives of order $m$. We will later also need to use the Sobolev $m$-norm, defined by $\|v\|_m := \sum_{i=0}^m |v|_i$. For $m=0$, $|v|_m = \|v\|_m = \|v\|$, so we have $\|u - \tilde{u}\| \leq C\, h^{k+1} |u|_{k+1}$.

The overall level of inaccuracy in approximating $u$ with $\widetilde{\tilde{u}}$ (and hence computing $\langle r, u \rangle$ from $\widetilde{\tilde{u}}$) can be bounded using the triangle inequality:
\[ \|u - \widetilde{\tilde{u}} \| \le \| u - \tilde{u} \| + \|\tilde{u} - \widetilde{\tilde{u}} \|. \]
To achieve a final error of $\epsilon \|r\|$ in computing $\langle r, u \rangle$ it is sufficient to achieve $\| u - \tilde{u} \| \le \epsilon/2$, $\|\tilde{u} - \widetilde{\tilde{u}} \| \le \epsilon/2$. Thus, by (\ref{eq:disc_error}), we can take a mesh such that
\begin{equation}\label{eq:classical_h}
h = O\left( \left(\frac{\epsilon}{|u|_{k+1}} \right)^{1/(k+1)}\right).
\end{equation}
Observe that $|u|_{k+1}$ might be initially unknown. In the case of the simple instance of the FEM discussed in the previous section, we had $|u|_{k+1} = \|f\|$, so this bound could be explicitly calculated. However, it can be nontrivial to estimate this quantity for more complicated BVPs.


\subsection{Classical complexity of the FEM}

The overall complexity of solving a BVP via the FEM is governed by the dimensionality of the problem being solved, the choice of finite element basis, and the desired accuracy criteria. These feed into the complexity of solving the required system of linear equations.

As the matrix $M$ is a Gramian matrix it is necessarily positive semidefinite. Also, the basis ${\phi_i}$ is almost universally chosen such that each basis vector only has support on a small number of finite elements, with the implication that $M$ is sparse, i.e.\ has $s = O(1)$ nonzero entries in each row. The most common choice of algorithm for inversion of matrices of this type (large, sparse, symmetric and positive semidefinite) is the \emph{conjugate gradient method}~\cite{shewchuk94} (for discussion in the context of the FEM, see \cite{axelsson01}, Sec. 1.3).
This method uses time $O(N s \sqrt{\kappa} \log 1/\epsilon_{CG})$ to solve a system $M \tilde{\mathbf{u}} = \tilde{\mathbf{f}}$ of $N$ linear equations, each containing at most $s$ terms, with condition number $\kappa = \|M\| \|M^{-1}\|$, up to accuracy $\epsilon_{CG}$ in the ``energy norm'' $\|\mathbf{x}\|_M := \sqrt{\mathbf{x}^T M \mathbf{x}}$.

We now estimate the values of each of the parameters in this complexity, first calculating the required size $N$. Let $\mathcal{P}$ be a basis for the space of polynomials of total degree $k$ in $d$ variables. To construct a basis for the space of piecewise degree-$k$ polynomials on the mesh, it is sufficient, for each finite element in the mesh, to include functions defined to be equal to a corresponding function in $\mathcal{P}$ on that finite element, and zero elsewhere. Then the total size of the basis is $N = O(h^{-d})$. Using~(\ref{eq:classical_h}), to achieve a final discretisation error of $\epsilon/2$ we can take
\[ N = O\left(\left(\frac{|u|_{k+1}}{\epsilon}\right)^{\frac{d}{k+1}}\right). \]
We next determine the required accuracy $\epsilon_{CG}$. Let $a$ be the inner product defining $M$, such that $M_{ij} = a(\phi_i,\phi_j)$. This inner product induces the energy norm (on functions) $\|u\|_E := \sqrt{a(u,u)}$. Use of this norm makes it easy to interpret the error from the conjugate gradient method, as one can readily calculate that
\begin{equation}
\label{eq:energymnorm}
\|\tilde{u} - \widetilde{\tilde{u}}\|_E = \| \tilde{\mathbf{u}} - \widetilde{\tilde{\mathbf{u}}}\|_M.
\end{equation}
In many important cases, such as elliptic PDEs, one can show that $a$ is {\em coercive}: there exists a universal constant $c$ such that $a(u,u) \ge c \|u\|^2$ for all $u$ (for further discussion on coercivity in PDEs, see~\cite{ern13}). It follows from coercivity of $a$ that $\| \tilde{u} - \widetilde{\tilde{u}}\| \le \sqrt{c} \| \tilde{u} - \widetilde{\tilde{u}}\|_E$. To achieve $\| \tilde{u} - \widetilde{\tilde{u}}\| \le \epsilon/2$ it is therefore sufficient to take $\epsilon_{CG} = O(\epsilon)$.

The scaling of $\kappa$, the condition number of $M$, with the size and shape of the mesh is discussed extensively in~\cite{bank89} and~\cite[Chapter 9]{brenner08}. Assume that $d \ge 2$ and that there exists a universal constant $C$ such that the basis functions $\phi_i$ satisfy
\begin{equation}\label{eq:basisnorm} C^{-1} h^{d-2} \|v\|_{L^\infty(T)} \le \sum_{\operatorname{supp}(\phi_i) \cap T \neq \emptyset} v_i^2 \le C h^{d-2} \|v\|_{L^\infty(T)} \end{equation}
for any function $v$ such that $v = \sum_i v_i \phi_i$, and any finite element $T$; this fixes the normalisation of the basis functions. Then, for a wide range of relatively regular meshes, the largest eigenvalue $\lambda_{\max}(M) = O(1)$ and the smallest eigenvalue $\lambda_{\min}(M) = \Omega(N^{-2/d})$, so $\kappa = O(N^{2/d})$. Finally, we have $s=O(1)$ by our assumption about the supports of the basis elements ${\phi_i}$. The overall complexity of the algorithm is thus
\[ O\left( \left(\frac{|u|_{k+1}}{\epsilon}\right)^{\frac{d+1}{k+1}} \log 1/\epsilon\right). \]
In many practical cases, however, \emph{preconditioning} is applied in order to reduce this scaling by improving the condition number. This can be seen as replacing the matrix $M$ with a matrix $M' = PM$ for some ``preconditioner'' $P$, and solving the new system of linear equations $M'\tilde{\mathbf{u}} = P \tilde{\mathbf{f}}$. A number of different preconditioners are known; one frequently used example in the case of the FEM is the sparse approximate inverse (SPAI) preconditioner. Although there is no guarantee that this preconditioner can improve the condition number in the worst case, experimental results suggest that it can be very effective in practice~\cite{benzi96,benzi99,li06,ping09}. If the condition number were reduced to the best possible scaling $O(1)$, we would obtain a ``best case'' runtime of the classical algorithm which is
\[
O\left( \left(\frac{|u|_{k+1}}{\epsilon}\right)^{\frac{d}{k+1}} \log 1/\epsilon\right).
\]
We remark that the preconditioned matrix $M'$ may no longer be symmetric; the dependence of the conjugate gradient method on the condition number $\kappa$ is quadratically worse for non-symmetric matrices, but as we have assumed that $\kappa=O(1)$ following preconditioning, this does not affect the complexity.

The best classical runtime following this approach is then found by optimising over allowed values of $k$. Observe that in either case, if $|u|_{k+1}$ and $d$ are fixed, this complexity is bounded by a polynomial in $1/\epsilon$.


\section{Solving the FEM with a quantum algorithm}
\label{sec:quantumfem}

The key step towards solving the FEM more quickly using a quantum computer is to replace the classical algorithm for solving the corresponding system of linear equations with a quantum algorithm. The fastest such algorithm known was recently presented by Childs, Kothari and Somma~\cite{childs15b}, improving previous algorithms of Harrow, Hassidim and Lloyd (HHL)~\cite{harrow09} and Ambainis~\cite{ambainis12b}.

\begin{thm}[Childs, Kothari and Somma~\cite{childs15b}]
\label{thm:cks}
Let $A$ be an $N \times N$ Hermitian matrix such that $\|A\|\|A^{-1}\| \le \kappa$, and $A$ has at most $s$ nonzero entries in each row. Assume there is an algorithm $\mathcal{P}_A$ which, on input $(r,i)$, outputs the location and value of the $i$'th nonzero entry in row $r$. Let $\mathbf{b}$ be an $N$-dimensional unit vector, and assume that there is an algorithm $\mathcal{P}_b$ which produces the corresponding state $\ket{b}$. Let
\[ \mathbf{x'} = A^{-1} \mathbf{b},\;\;\;\; \ket{x} = \frac{\mathbf{x'}}{\|\mathbf{x'}\|}. \]
Then there is a quantum algorithm which produces the state $\ket{x}$ up to accuracy $\epsilon$ in $\ell_2$ norm, with bounded probability of failure, and makes
\[ O\left(s\kappa \poly(\log(s\kappa/\epsilon)) \right) \]
uses of $\mathcal{P}_A$ and $\mathcal{P}_b$. The runtime is the same up to a $\poly(\log N)$ factor.
\end{thm}

We will also need to approximate the Euclidean norm of the solution, $\|\mathbf{x'}\|$. The most efficient approach known to achieve this appears to be based on the original HHL algorithm. The number of uses of $\mathcal{P}_A$ required to estimate $\|\mathbf{x'}\|$ up to accuracy $\epsilon \|\mathbf{x'}\|$ can be shown to be
\[ O((s \kappa^2/\epsilon) \poly \log(s\kappa/\epsilon)); \]
the number of uses of $\mathcal{P}_b$ required is $O(\kappa/\epsilon)$. Assuming that $\mathcal{P}_A$ and $\mathcal{P}_b$ can each be implemented in time $\poly(\log N)$, the runtime of the algorithm is 
\[ O((s \kappa^2/\epsilon) \poly \log(Ns\kappa/\epsilon)). \]
As we were unable to find statements of these bounds in the literature, we sketch the argument behind them in Appendix~\ref{sec:hhlnorm}.

Here we will apply these results to the linear system $M \tilde{\mathbf{u}} = \tilde{\mathbf{f}}$. We see from the above bounds that the complexity of the overall quantum algorithm for solving the FEM is determined by the following parameters:
\begin{enumerate}
\item The complexities of the algorithms $\mathcal{P}_M$ and $\mathcal{P}_{\tilde{f}}$, which respectively determine elements of $M$ and (approximately) produce $\ket{\tilde{f}}$;
\item The condition number $\kappa$ and sparsity $s$ of the matrix $M$;
\item The complexity of determining some quantity of interest given a state which approximates $\ket{\tilde{u}}$.
\end{enumerate}
These quantities will depend in turn on the desired accuracy of the output. We now investigate each of them.

Note that most of the algorithms we use will have some arbitrarily small, but non-zero, probability of failure. We assume throughout that failure probabilities have been made sufficiently low that they can be disregarded.


\subsection{Preparing the input}

The purpose of this section is to discuss the time to prepare the input state $\ket{\tilde{f}}$ (i.e.\ the complexity of the subroutine $\mathcal{P}_{\tilde{f}}$ required for the QLE algorithm). To achieve an efficient algorithm overall, we would like to be able to prepare $\ket{\tilde{f}}$ in time $\poly(\log N)$. Rather than rely on a quantum RAM to provide $\ket{\tilde{f}}$, we instead refer to a scheme introduced by Zalka~\cite{zalka98}, and independently rediscovered by both Grover and Rudolph~\cite{grover02} and Kaye and Mosca~\cite{kaye04}.

The scheme can be used to produce a real quantum state $\ket{\psi}$ of $n$ qubits in time polynomial in $n$, given the ability to compute the weights
\[ W_x := \sum_{y \in \{0,1\}^{n-k}} |\langle xy | \psi \rangle|^2 \]
for arbitrary $k=1,\dots,n$ and arbitrary $x \in \{0,1\}^k$ in time $\poly(n)$, as well as the ability to determine the sign of $\langle x | \psi \rangle$ for arbitrary $x$ in time $\poly(n)$.

To approximately produce $\ket{\psi}$ up to a high level of accuracy (e.g.\ $O(2^{-n})$) in time polynomial in $n$, it is actually sufficient to be able to approximately compute each weight $W_x$ up to accuracy $\epsilon$ in time $O(\log 1/\epsilon)$, for arbitrary $\epsilon$. We sketch the argument as follows. The algorithm of~\cite{zalka98,grover02,kaye04} is designed to produce a state $\ket{\psi'}$ with non-negative amplitudes in the computational basis, such that $\ip{x}{\psi'} = |\ip{x}{\psi}| = \sqrt{W_x}$ for all $x \in \{0,1\}^n$, and then flips the signs of amplitudes as required. To produce $\ket{\psi'}$ the algorithm expresses $W_x$, for each $x \in \{0,1\}^n$, as a telescoping product
\[ W_x = W_{x_1}\times \frac{W_{x_1 x_2}}{W_{x_1}} \times \frac{W_{x_1 x_2 x_3}}{W_{x_1 x_2}} \times \dots \times \frac{W_x}{W_{x_1\dots x_{n-1}}}, \]
computes each fraction in turn (in superposition), and uses this to set $\ip{x}{\psi'}$. If the goal is to produce $\ket{\psi}$ up to accuracy $\epsilon$ in $\ell_2$ norm, from the inequality $(|\ip{x}{\psi}| - |\ip{x}{\psi'}|)^2 \le |\ip{x}{\psi}^2 - \ip{x}{\psi'}^2|$ it is sufficient to approximate each weight $W_x$, $x \in \{0,1\}^n$, up to additive error $O(\epsilon^2 / 2^n)$. So the product can be truncated at the point $i$ where the weight $W_{x_1\dots x_i} = O(\epsilon^2/2^n)$, because any subsequent multiplications can only decrease $W_x$, and weights below this size can be ignored.

If the algorithm does not compute weights $W_x$, $W_y$ in some fraction $W_x/W_y$ exactly, but instead computes approximations $\widetilde{W_x}$ and $\widetilde{W_y}$ such that $|\widetilde{W_x} - W_x| \le \gamma W_x$ and $|\widetilde{W_y} - W_y| \le \gamma W_y$ for some $\gamma$, then $| \widetilde{W_x}/\widetilde{W_y} - W_x/W_y| = O(\gamma W_x / W_y)$. As we have assumed that $W_x = \Omega(\epsilon^2/2^n)$ for all $k$-bit strings $x$ for which we compute $W_x$ ($1 \le k \le n$), it is sufficient to approximate each weight $W_x$ up to additive accuracy $O(\epsilon^2 / (n2^n))$ for each fraction to be accurate up to a multiplicative error of $O(\epsilon^2/(n2^n))$ and hence the overall product of weights to be accurate up to an additive error $O(\epsilon^2 / 2^n)$. From the assumption about the complexity of the algorithm for approximately computing $W_x$, we can achieve this level of accuracy in $\poly(n, \log 1/\epsilon)$ time.

In the case of the FEM, the weights $W_x$ correspond to quantities of the form
\[ S(a,b) := \sum_{i=a}^b \left( \int_\Omega \phi_i(\mathbf{x}) f(\mathbf{x}) d\mathbf{x} \right)^2, \]
%
where $\mathbf{x} \in \mathbb{R}^d$ and $a$ and $b$ are integers. Expressions of this form can be computed (either exactly or approximately) for many functions $f$ of interest. For example, consider the 1-dimensional setting discussed in Section \ref{sec:femintro}. If $f$ is a polynomial, then the integral can be easily calculated, and corresponds to a polynomial in $x_{i-1}$, $x_i$ and $x_{i+1}$. If the finite elements are regularly spaced, so $x_i = ih$ for some $h$, the entire sum $S(a,b)$ is a polynomial in $a$ and $b$ which can be explicitly calculated for any $a$ and $b$.

For a choice of polynomial basis of degree $k$ (i.e.\ where the $(k+1)^{th}$ derivative $\phi^{(k+1)}_i$ = 0), then from Darboux's formula one has that
\[ \int_\Omega \phi_i(\mathbf{x}) f(\mathbf{x}) d\mathbf{x} = \sum_{j=1}^{k} (-1)^{j} \phi_i^{(j)}(\mathbf{x}) \underbrace{\int \cdots \int}_{j+1 \text{ times}} f(\mathbf{x}) d\mathbf{x}.\]
So, once the basis is specified, computing individual amplitudes is only as difficult as integrating the function $f(\mathbf{x})$. However, the state-production algorithm requires the computation of weights which depend on up to $N = 2^n$ squared amplitudes. To obtain an efficient algorithm, it is therefore necessary to find a more concise expression for these sums.

As discussed above, this can be achieved when $f$ is a polynomial and the finite element mesh is suitably regular. This includes some physically interesting cases; even a constant function $f$ can be of interest. An efficiently computable expression for $S(a,b)$ can also be obtained when $f$ is only supported on a few basis elements. However, it appears challenging to compute this quantity efficiently for more general functions $f$. Indeed, see Section~\ref{sec:oracular} for an argument that this should not be achievable in general.

For simplicity in the subsequent bounds, we henceforth assume that the state $\ket{\tilde{f}}$ can be produced perfectly in time $\poly(\log N)$.


\subsection{Solving the system of linear equations}

Let $M$ be the matrix defined by $M_{ij} = a(\phi_i,\phi_j)$. Recall from Theorem \ref{thm:cks} that the quantum algorithm assumes that it has access to an algorithm $\mathcal{P}_M$ which, on input $(r,i)$, outputs the location and value of the $i$'th nonzero entry in row $r$ (or ``not found'' if there are fewer than $i$ nonzero entries). If the finite element mesh is suitably regular, $\mathcal{P}_M$ is easy to implement. For instance, consider the set of $n$ piecewise linear functions on $[0,1]$ defined in Section \ref{sec:femintro}. Then $M$ is a tridiagonal matrix whose diagonal elements are equal to $2/h$, and whose off-diagonal elements are equal to $-1/h$. Hence for $r > 1$,
\[ \mathcal{P}_M(r,i) = \begin{cases} (r-1,-1/h) & \text{ if $i=1$ }\\ (r,2/h) & \text{ if $i=2$}\\ (r+1,-1/h) & \text{ if $i=3$,}\\ \text{not found} & \text{ otherwise.} \end{cases} \]
More generally, it will be possible to implement $\mathcal{P}_M$ efficiently if there is an efficient procedure for mapping an index to a finite element, and for listing the neighbouring elements for a given element. This will be the case, for example, when the finite element mesh is a regular triangulation of a polygon. For discussion on automated mesh generation and indexing schemes, see both~\cite{haber81} and Sec.\ 5.1 in~\cite{axelsson01}.

When solving the system of linear equations, inaccuracies in the prepared state $\ket{\tilde{f}}$ will translate into inaccuracies in the output state $\ket{\tilde{u}}$. Let $\widetilde{\ket{\tilde{f}}}$ be the approximate state that was actually prepared. Then the state produced after applying the QLE algorithm is (approximately)
\[ \frac{M^{-1}\widetilde{\ket{\tilde{f}}}}{\|M^{-1}\widetilde{\ket{\tilde{f}}}\|}. \]
If $\ket{\tilde{f}}$ is prepared up to accuracy $\epsilon$ in $\ell_2$ norm, then the inaccuracy of the output state in $\ell_2$ norm is
\[ \left\| \frac{M^{-1}\ket{\tilde{f}}}{\|M^{-1}\ket{\tilde{f}}\|} - \frac{M^{-1}\widetilde{\ket{\tilde{f}}}}{\|M^{-1}\widetilde{\ket{\tilde{f}}}\|} \right\|. \]
Writing $\widetilde{\ket{\tilde{f}}} = \ket{\tilde{f}} + \ket{\epsilon}$ for some vector $\ket{\epsilon}$ such that $\|\ket{\epsilon}\| = \epsilon$, this quantity is equal to
\begin{align*}
& \left\| \frac{M^{-1}\ket{\tilde{f}}(\|M^{-1} \widetilde{\ket{\tilde{f}}}\| - \|M^{-1}\ket{\tilde{f}}\|)}{\|M^{-1}\ket{\tilde{f}}\|\|M^{-1}\widetilde{\ket{\tilde{f}}}\|} - \frac{M^{-1} \ket{\epsilon}}{\|M^{-1} \widetilde{\ket{\tilde{f}}}\|} \right\| \\
&\le \frac{| \|M^{-1} \widetilde{\ket{\tilde{f}}}\| - \|M^{-1}\ket{\tilde{f}}\| | }{\|M^{-1} \widetilde{\ket{\tilde{f}}}\|} + \frac{\|M^{-1} \ket{\epsilon}\|}{\|M^{-1} \widetilde{\ket{\tilde{f}}}\|}\\
&\le 2 \frac{\|M^{-1}\ket{\epsilon}\|}{\|M^{-1} \widetilde{\ket{\tilde{f}}}\|}\\
&\le 2\epsilon \kappa,
\end{align*}
by the triangle inequality, the reverse triangle inequality, and the definition of the condition number $\kappa$. We therefore see that, if preconditioning is not applied to the matrix $M$ to reduce $\kappa$, it is necessary for $\ket{\tilde{f}}$ to be prepared up to accuracy $O(N^{-2/d} \epsilon)$. (Note that this is not an issue if we can produce $\ket{\tilde{f}}$ exactly, as for some examples discussed in the previous section.)

Clader, Jacobs and Sprouse~\cite{clader13} showed that the sparse approximate inverse (SPAI) preconditioner can be used within the overall framework of the QLE algorithm. Preconditioning replaces $M$ with $M' = PM$ for some matrix $P$, to obtain the corresponding linear system $PM \tilde{\mathbf{u}} = P \tilde{\mathbf{f}}$. In the SPAI preconditioner, $P$ is chosen such that $P \approx M^{-1}$ and also that $P$ is sparse. The sparsity desired is a parameter of the algorithm; although one has no guarantees that either $P$ or $PM$ will be sparse while $PM$ achieves a low condition number, in practice this is often the case. The structure of the SPAI is designed such that queries to entries of $PM$ can be computed from queries to $M$ with a modest overhead~\cite{clader13}.

If preconditioning is used, we no longer need to prepare the initial state $\ket{\tilde{f}}$, but a state proportional to $P\ket{\tilde{f}}$. Note that preparing the input $P\tilde{\mathbf{f}}$ in the classical case requires only multiplication of a vector by a sparse matrix, which is computationally cheap compared to matrix inversion. As such, it is typically neglected when considering the classical computational complexity. However, the situation is more complicated in the quantum setting.

The most straightforward way to prepare $P\ket{\tilde{f}}$ is to construct $\ket{\tilde{f}}$ and then attempt to apply the (non-unitary) operation $P$. There are several known approaches which can be used to achieve this probabilistically. One elegant example is a simple special case of the ``Chebyshev'' approach of~\cite[Section 4]{childs15b}. This work uses a quantum walk to apply $n$'th order Chebyshev polynomials $T_n(P)$ in an arbitrary $s$-sparse Hermitian matrix $P$. (If $P$ is not Hermitian, a standard trick~\cite{harrow09} can be used to express it as a submatrix of a Hermitian matrix.) As the first Chebyshev polynomial $T_1$ is simply $T_1(x) = x$, this allows $P$ itself to be implemented. If the subroutine of~\cite{childs15b} succeeds when applied to a state $\ket{\psi}$, then $\ket{\psi}$ is (exactly) mapped to $P\ket{\psi}/\|P\ket{\psi}\|$. The success probability is at least
\[ \frac{\| P \ket{\psi} \|^2 }{s^2 \|P\|_{\max}^2} \ge \frac{1}{\kappa(P)^2 s^2}, \]
where $\|P\|_{\max} = \max_{i,j} |P_{ij}|$ and we use $\|P\|_{\max} \le \|P\|$. Using amplitude amplification, the failure probability can be made at most $\delta$, for arbitrarily small $\delta > 0$, with $O(1/(\kappa(P)s))$ repetitions. Each repetition requires time polylogarithmic in $N$, $\kappa(P)$ and $s$.

Combining all these considerations, we see that if preconditioning is used, the complexity of the quantum algorithm will depend on a number of different parameters, each of which may be hard to estimate in advance. These are: the condition number of $PM$; the complexity of computing entries of $P$; the sparsity of $P$; and the condition number of $P$. Here, in order to give a ``best case'' comparison of the preconditioned quantum algorithm with the classical algorithm, we make the optimistic assumption that preconditioning is optimal (i.e.\ $\kappa(PM) = O(1)$), and that taking all of these additional sources of complexity into account multiplies the runtime by only a $\poly(\log(N))$ factor.


\subsection{Measuring the output}
\label{sec:output}

By running the QLE algorithm, we obtain an output state $\ket{\widetilde{\tilde{u}}}$ which approximates the normalised state
\[ \ket{\tilde{u}} = \frac{\sum_i \tilde{\mathbf{u}}_i \ket{i}}{\sqrt{\sum_i \tilde{\mathbf{u}}_i^2}}, \]
where we associate each basis state $\ket{i}$ with the basis function $\phi_i$. Given copies of $\ket{\tilde{u}}$, we can carry out measurements to extract information about $u$. One example is the prototypical problem we consider here, approximating the $L^2$ inner product $\langle u,r \rangle$ between $u$ and a fixed function $r$. This can be achieved by approximately computing the inner product $\ip{\tilde{u}}{r}$ between $\ket{\tilde{u}}$ and the state $\ket{r}$ defined by
\begin{equation} \label{eq:r} \ket{r} = \frac{1}{(\sum_i \langle \phi_i, r \rangle^2)^{1/2}} \sum_i \langle \phi_i,r\rangle  \ket{i} \end{equation}
for some function $r$; then $\ip{\tilde{u}}{r}$ is the $L^2$ inner product between $\tilde{u}$ and $r$, up to an overall scaling factor. $\ket{r}$ can be produced using techniques described in the previous section. Some interesting cases are particularly simple: for example, taking $r$ to be uniform on a region, $\ip{\tilde{u}}{r}$ gives the average of $\tilde{u}$ over that region.

This inner product can be estimated using a procedure known as the Hadamard test~\cite{aharonov09a}, a subroutine whose output is a $\pm1$-valued random variable with expectation $\ip{\tilde{u}}{r}$. By applying amplitude estimation~\cite{brassard02} to approximately compute this expectation, $\langle \tilde{u} | r \rangle$ can be estimated up to accuracy $\epsilon$ with $O(1/\epsilon)$ uses of algorithms to produce the states $\ket{\tilde{u}}$ and $\ket{r}$. A related approach was used by Clader, Jacobs and Sprouse~\cite{clader13} to compute an electromagnetic scattering cross-section, which corresponds to a quantity of the form $|\langle \tilde{u} | r \rangle|^2$. This can be approximately computed using the swap test~\cite{buhrman01}, a subroutine which, given two states $\ket{\psi}$, $\ket{\psi'}$, outputs ``same'' with probability $\frac{1}{2} + \frac{1}{2} |\langle \psi | \psi' \rangle|^2$, and ``different'' otherwise.

We remark that more complicated properties of $u$ seem to be more problematic to compute directly from the state $\ket{\tilde{u}}$, due to the non-orthogonality of the basis $\{\phi_i\}$. For example, one common use of the QLE algorithm is to determine similarity of solutions to sets of linear equations by using the swap or Hadamard tests to compare them~\cite{harrow09,ambainis12b}. Consider two states $\ket{a}$, $\ket{b}$ corresponding to functions $a = \sum_i \mathbf{a}_i \phi_i$, $b = \sum_i \mathbf{b}_i \phi_i$. Then
\[ \ip{a}{b} \propto \sum_i \mathbf{a}_i \mathbf{b}_i \]
while a sensible measure of similarity of the functions $a$ and $b$ is the inner product
\[ \int_\Omega a(\mathbf{x}) b(\mathbf{x}) d\mathbf{x} = \sum_{i,j} \mathbf{a}_i \mathbf{b}_j \int_{\Omega} \phi_i(\mathbf{x}) \phi_j(\mathbf{x}) d\mathbf{x}. \]
One would hope for this to be approximately proportional to $\sum_i \mathbf{a}_i \mathbf{b}_i$. However, although $\phi_i$ and $\phi_j$ do not have overlapping support for most pairs $i \neq j$, there are still enough such pairs where this overlap is nonzero that the integral can sometimes be a poor approximation.


\subsection{Overall complexity}
\label{sec:overall}

The total complexity of the quantum algorithm for solving an FEM problem is found by combining the complexities of all of the above pieces.

Assume that we would like to compute $R := \int_\Omega r(\mathbf{x}) u(\mathbf{x}) d\mathbf{x}$ for some $r:\Omega \rightarrow \mathbb{R}$ up to additive error $\epsilon \|r\|$. Write $\alpha = (\sum_i \langle \phi_i,r \rangle^2)^{1/2}$. The quantum algorithm will perform the following steps by applying the QLE algorithm to the system of linear equations $M \tilde{\mathbf{u}} = \tilde{\mathbf{f}}$:
\begin{enumerate}
\item Estimate $\|\tilde{\mathbf{u}}\|$ up to an additive term $\epsilon_N$. Let $\widetilde{N}$ be the estimate.
\item Use the QLE algorithm to produce copies of $\ket{\widetilde{\tilde{u}}}$, an approximation to $\ket{\tilde{u}}$. Use these to estimate $\ip{r}{\widetilde{\tilde{u}}}$ up to an additive term $\epsilon_{\text{out}}$. Let $\widetilde{R}$ be the estimate.
\item Output $\alpha \widetilde{N} \widetilde{R}$ as an estimate of $R$.
\end{enumerate}
We can bound the overall error as follows. Let $\epsilon_L$ be the inaccuracy, in $\ell_2$ norm, in solving the system of linear equations in step (2), i.e.\ $\epsilon_L = \| \ket{\widetilde{\tilde{u}}} - \ket{\tilde{u}}\|$. This encompasses any error in producing the initial state $\ket{\tilde{f}}$, as well as inaccuracy arising from the QLE algorithm itself (although recall that we have in fact assumed that we can produce $\ket{\tilde{f}}$ perfectly). Then
\begin{eqnarray*}
\widetilde{R} &=& \ip{r}{\widetilde{\tilde{u}}} + \epsilon_{\text{out}}\\
&=& \ip{r}{\tilde{u}} + \langle r | (\ket{ \widetilde{\tilde{u}}} - \ket{\tilde{u}}) + \epsilon_{\text{out}}\\
&=& \ip{r}{\tilde{u}} + \epsilon_L' + \epsilon_{\text{out}}
\end{eqnarray*}
for some $\epsilon_L'$, where $|\epsilon_L'| \le \epsilon_L$ by Cauchy-Schwarz. So
\[
\widetilde{R} = \frac{\sum_i \tilde{\mathbf{u}}_i \langle \phi_i,r \rangle}{\|\tilde{\mathbf{u}}\| \left(\sum_i \langle \phi_i,r \rangle^2 \right)^{1/2}} + \epsilon_L' + \epsilon_{\text{out}} = \frac{\langle \tilde{u}, r \rangle}{\alpha \|\tilde{\mathbf{u}}\|} +\epsilon'_L + \epsilon_{\text{out}}
\]
using the definition of $\ket{r}$ from (\ref{eq:r}) and of $\ket{\tilde{u}}$ as a normalised version of $\tilde{\mathbf{u}}$. Writing $\widetilde{N} = \|\tilde{\mathbf{u}}\| + \epsilon_N$, we have
\[
\alpha \widetilde{N} \widetilde{R} = \langle \tilde{u}, r \rangle\left(1+\frac{\epsilon_N}{\|\tilde{\mathbf{u}}\|}\right) + \alpha(\|\tilde{\mathbf{u}}\|+\epsilon_N) (\epsilon_L' + \epsilon_{\text{out}}).
\]
The analysis of the remaining term $\langle \tilde{u}, r \rangle$ is now similar to the classical setting:
\[
\langle \tilde{u}, r \rangle = \langle u, r \rangle + \langle \tilde{u} - u,r \rangle = \langle u, r \rangle + \epsilon_D',
\]
where
\[ |\epsilon_D'| \le \|r\| \|\tilde{u} - u\| =: \|r\| \epsilon_D \]
by Cauchy-Schwarz. Combining all the terms together, we have
\begin{multline*} \alpha \widetilde{N} \widetilde{R} - R = \\ \epsilon'_D \left(1 + \frac{\epsilon_N}{\|\tilde{\mathbf{u}}\|} \right) + \frac{\langle u,r\rangle \epsilon_N}{\|\tilde{\mathbf{u}}\|} + \alpha(\|\tilde{\mathbf{u}}\|+\epsilon_N) (\epsilon_L' + \epsilon_{\text{out}}).
\end{multline*}
We finally use $\langle u,r\rangle \le \|u\| \|r\|$ to obtain
\begin{multline*} \alpha \widetilde{N} \widetilde{R} - R = \\ \epsilon'_D \left(1 + \frac{\epsilon_N}{\|\tilde{\mathbf{u}}\|} \right) + \frac{\|u\| \|r\| \epsilon'_N}{\|\tilde{\mathbf{u}}\|} + \alpha(\|\tilde{\mathbf{u}}\|+\epsilon_N) (\epsilon_L' + \epsilon_{\text{out}})
\end{multline*}
for some $\epsilon'_N$ such that $|\epsilon'_N| \le |\epsilon_N|$. To achieve overall accuracy $\epsilon \|r\|$ it is sufficient for each term in this expression to be upper-bounded by $\epsilon\|r\|/3$, which follows from
\begin{eqnarray*}
\epsilon_D &=& O(\epsilon),\\
\epsilon_N &=& O(\min \{ \|\widetilde{\mathbf{u}} \|, \epsilon \|\tilde{\mathbf{u}}\|/\|u\| \}),\\
\epsilon_L, \epsilon_{\text{out}} &=& O(\epsilon\|r\|/(\alpha \|\tilde{\mathbf{u}} \|)).
\end{eqnarray*}
Assume for simplicity in the final bound that $\epsilon \le \|u\|$; then the second condition becomes $\epsilon_N = O(\epsilon \|\tilde{\mathbf{u}}\|/\|u\|)$. We now calculate the complexity of achieving these accuracies.

Using the discretisation error bound (\ref{eq:disc_error}), we have $\epsilon_D \leq C\, h^{k+1} |u|_{k+1}$ for some universal constant $C$. We can therefore take
\[ h = O\left( \left(\frac{\epsilon}{|u|_{k+1}} \right)^{\frac{1}{k+1}}\right). \]
As in the classical case, this choice of $h$ corresponds to solving a system of
\[ N = O\left( \left(\frac{|u|_{k+1}}{\epsilon}\right)^{\frac{d}{k+1}}\right) \]
linear equations. For any $\delta > 0$, as discussed in Section \ref{sec:quantumfem} and Appendix \ref{sec:hhlnorm}, $\|\tilde{\mathbf{u}}\|$ can be approximated up to accuracy $\delta\|\tilde{\mathbf{u}}\|$ in time $O((s \kappa^2 / \delta)\poly \log(Ns\kappa/\delta))$ using the HHL algorithm, recalling that $s$ and $\kappa$ are the sparsity and condition number of $M$, respectively. Inserting $\delta = \epsilon/\|u\|$ and the bound on $N$, this part requires time
\[ O\left(\frac{s \kappa^2\|u\|}{\epsilon} \poly(\log(s\kappa\|u\| |u|_{k+1}/\epsilon)) \right) \]
We can also put an upper bound on $\epsilon_L$ and $\epsilon_{\text{out}}$ by upper-bounding $\alpha/\|r\|$ and $\| \tilde{\mathbf{u}} \|$. In the former case, it holds that
\[ \frac{\alpha}{\|r\|} = O(h \sqrt{s}); \]
we prove this technical claim in Appendix \ref{sec:techbounds}. In the latter case,
\[ \| \tilde{\mathbf{u}} \| = O(\|\tilde{\mathbf{u}}\|_M/h) = O(\|\tilde{u}\|_E/h) = O(\|u\|_1/h). \]
The first two equalities follow from $\lambda_{\min}(M) = \Omega(N^{-2/d}) = \Omega(h^2)$~\cite{bank89,brenner08} and the equivalence between $\|\tilde{\mathbf{u}}\|_M$ and $\|\tilde{u}\|_E$ (see (\ref{eq:energymnorm})). The third follows from $\|\tilde{u}\|_E = O(\|\tilde{u}\|_1)$, where $\|\cdot\|_1$ is the Sobolev 1-norm, which is a consequence of the inner product $a(\cdot,\cdot)$ defining the energy norm corresponding to an underlying elliptic PDE~\cite{brenner08}, and the bound (\ref{eq:disc_error}). Combining these bounds, the requirement on $\epsilon_L$ and $\epsilon_{\text{out}}$ can be rewritten as
\[ \epsilon_L, \epsilon_{\text{out}} = O\left(\frac{\epsilon}{\sqrt{s} \|u\|_1}\right). \]
To achieve accuracy $\epsilon_{\text{out}}$ using the Hadamard test and amplitude estimation requires $O(1/\epsilon_{\text{out}})$ uses of the QLE algorithm, each of which runs in time $O(s \kappa \poly(\log(N s \kappa/\epsilon_L)))$ by Theorem \ref{thm:cks}. Inserting the bounds on $\epsilon_L$, $\epsilon_{\text{out}}$ gives a complexity for part (2) which is
\[ O \left( \frac{\sqrt{s} \kappa \|u\|_1}{\epsilon} \poly(\log(s \kappa \|u\|_1 |u|_{k+1}/\epsilon))  \right) \]
Combining these bounds, we obtain an overall runtime of
\[ O \left(\frac{s\kappa^2\|u\| + \sqrt{s}\kappa \|u\|_1}{\epsilon} \poly(\log(s \kappa \|u\|_1 |u|_{k+1}/\epsilon)) \right). \]
In fixed spatial dimension, $s=O(1)$, and if preconditioning is not used, $\kappa = O(N^{2/d}) = O((|u|_{k+1}/\epsilon)^{2/(k+1)})$. Inserting these values, we obtain a bound of
\[ \widetilde{O} \left( \frac{\|u\| |u|_{k+1}^{\frac{4}{k+1}}}{\epsilon^{\frac{k+5}{k+1}}} + \frac{\|u\|_1 |u|_{k+1}^{\frac{2}{k+1}}}{\epsilon^{\frac{k+3}{k+1}}} \right), \]
where the $\widetilde{O}$ notation hides polylogarithmic factors. On the other hand, if we assume that optimal preconditioning has been applied, so $\kappa$ reduces to $O(1)$, we would obtain an overall bound of just
\[ \widetilde{O} \left(\frac{\|u\|_1}{\epsilon} \right). \]
These runtimes should be compared with the corresponding runtimes of the classical algorithm:
\[ \widetilde{O}\left( \left(\frac{|u|_{k+1}}{\epsilon}\right)^{\frac{d+1}{k+1}}\right)\;\; \text{ and } \;\; \widetilde{O}\left( \left(\frac{|u|_{k+1}}{\epsilon}\right)^{\frac{d}{k+1}}\right), \]
respectively. First, note that if preconditioning is used, the dependence on $|u|_{k+1}$ in the quantum algorithm's runtime is significantly milder than for the classical algorithm, being only polylogarithmic. Even if preconditioning is not used, for large enough $d$ the dependence is polynomially better.

Perhaps more importantly, observe that in the term dependent on $\epsilon$ in the algorithms' runtimes, the quantum algorithm's runtime no longer depends on the dimension $d$. This holds whether or not preconditioning is used. Thus the quantum algorithm will achieve a large speedup when $\epsilon$ is small and $d$ is large. (Note that we cannot quite call this an exponential quantum speedup with respect to $d$: the runtime of the quantum algorithm also depends on a term $C$ in (\ref{eq:disc_error}) which is constant for a fixed dimension and family of meshes, but has unbounded dependence on $d$.)

As a demonstrative example of the speedup (or otherwise) expected for the quantum algorithm, consider the case of solving a BVP in four dimensions (three spatial and one temporal, say), using piecewise linear basis functions. Then, the classical runtimes both without preconditioning and with optimal preconditioning are
\[ \widetilde{O}\left( \left(\frac{|u|_{2}}{\epsilon}\right)^{\frac{5}{2}}\right)\;\; \text{ and } \;\; \widetilde{O}\left( \left(\frac{|u|_{2}}{\epsilon}\right)^{2}\right), \]
respectively. The analogous quantum runtimes are
\[ \widetilde{O} \left( \frac{\|u\| |u|_2^2}{\epsilon^3} + \frac{\|u\|_1 |u|_2}{\epsilon^2} \right)\;\; \text{ and } \;\; \widetilde{O} \left(\frac{\|u\|_1}{\epsilon} \right). \]
In this case, lack of preconditioning leads to a quantum algorithm which might or might not outperform the classical algorithm, depending on the relative sizes of $\epsilon$, $\|u\|$, $\|u\|_1$ and $|u|_2$. In the optimally preconditioned case, the quantum algorithm both scales better with accuracy and has a less stringent condition on the solution smoothness.


\section{Quantum lower bounds}
\label{sec:quantumlower}

We have seen that the QLE algorithm can be used to obtain polynomial quantum speedups over the best known classical algorithms for the FEM. We now argue that, in the physically realistic setting of fixed dimension and smooth solutions, a polynomial quantum speedup is the largest speedup one can expect. We first discuss the general question of putting lower bounds on the complexity of algorithms based on a QLE subroutine.


\subsection{A general quantum lower bound}
\label{sec:lower1}

We observe from Theorem \ref{thm:cks} and the discussion in Section \ref{sec:output} that producing the quantum state $\ket{x} \propto A^{-1} \ket{b}$ for some well-conditioned, sparse matrix $A$ can be achieved in time $\poly\log(1/\epsilon)$, while the apparently simpler task of approximating some natural properties of $\ket{x}$ uses time $O(1/\epsilon)$. It is therefore natural to suspect that the runtime of this component could be improved substantially, e.g.\ to $\poly\log(1/\epsilon)$. However, it was shown by Harrow, Hassidim and Lloyd~\cite{harrow09} that the existence of a quantum algorithm with this scaling for approximating some very simple properties of $\ket{x}$ would imply the complexity-theoretic consequence BQP=PP, which is considered highly unlikely (implying, for example, that quantum computers could efficiently solve NP-complete problems).

As well as this complexity-theoretic argument, we now give an argument based on ideas from query complexity which lower bounds the runtime of any algorithm which approximates some function of the output of the QLE algorithm, without making use of the internal structure of the algorithm. This encompasses all the uses of QLE for the FEM discussed in Section \ref{sec:output}.

We adapt a standard technique of Bennett et al.~\cite{bennett97}. Consider an algorithm which has access to a unitary subroutine $\mathcal{A}_{\psi}$, parametrised by an unknown state $\ket{\psi}$, such that $\mathcal{A}_\psi$ maps $\ket{0}$ to $\ket{\psi}$. The algorithm may also have access to the inverse subroutine $\mathcal{A}_\psi^{-1}$. The algorithm does not know anything about how $\mathcal{A}_\psi$ is implemented and uses it as a ``black box''. It aims to estimate some property of $\ket{\psi}$. In the context of the FEM, we think of $\mathcal{A}_{\psi}$ as the QLE algorithm, where $\ket{\psi}$ is the output state corresponding to the approximate solution of the desired BVP. We assume that the algorithm only makes use of the QLE subroutine for one instance, i.e.\ it uses $\mathcal{A}_\psi$ throughout, rather than $\mathcal{A}_{\psi'}$ for some $\ket{\psi'} \neq \ket{\psi}$; relaxing this assumption would only make the problem harder.

For notational simplicity, we also assume in the proof that the overall algorithm does not use $\mathcal{A}_\psi^{-1}$ and does not use any ancilla qubits. (These assumptions can easily be relaxed without changing the conclusions.) Further assume that the overall algorithm makes $T$ uses of $\mathcal{A}_\psi$, interspersed with arbitrary unitary operators $U_1,\dots,U_{T+1}$. Let $\ket{\phi}$ be such that $\| \ket{\psi} - \ket{\phi}\| \le \epsilon$, and such that the output of the algorithm should be different when using $\mathcal{A}_\phi$ rather than $\mathcal{A}_\psi$. Finally let $\ket{\eta}_{\psi,t}$ be the state of the overall algorithm after $t$ uses of $\mathcal{A}_\psi$. Then

\begin{align*}
& \| \ket{\eta}_{\psi,T} - \ket{\eta}_{\phi,T} \|\\
&= \| U_{T+1} \mathcal{A}_\psi U_T \dots \mathcal{A}_\psi U_1 \ket{0} - U_{T+1} \mathcal{A}_\phi U_T \dots \mathcal{A}_\phi U_1 \ket{0} \| \\
 &= \| \mathcal{A}_\psi U_T \dots \mathcal{A}_\psi U_1 \ket{0} - \mathcal{A}_\phi U_T \dots \mathcal{A}_\phi U_1 \ket{0} \| \\
 &\le \| \mathcal{A}_\psi U_T \mathcal{A}_\psi U_{T-1} \mathcal{A}_\psi \dots \mathcal{A}_\psi U_1 \ket{0} \\
 &\hspace{0.5cm}- \mathcal{A}_\psi U_T \mathcal{A}_\phi U_{T-1} \mathcal{A}_\phi \dots \mathcal{A}_\phi U_1 \ket{0} \| \\
 &+ \| \mathcal{A}_\psi U_T \mathcal{A}_\phi U_{T-1} \mathcal{A}_\phi \dots \mathcal{A}_\phi U_1 \ket{0} - \mathcal{A}_\phi U_T \dots \mathcal{A}_\phi U_1 \ket{0} \| \\
 &\le \| \mathcal{A}_\psi U_{T-1} \mathcal{A}_\psi \dots \mathcal{A}_\psi U_1 \ket{0} - \mathcal{A}_\phi U_{T-1} \mathcal{A}_\phi \dots \mathcal{A}_\phi U_1 \ket{0} \|\\
 &\hspace{0.5cm} + \| \mathcal{A}_\psi - \mathcal{A}_\phi \|.
\end{align*}
The first inequality is the triangle inequality, while the second uses the fact that unitaries do not change the Euclidean distance. As the algorithm does not use any information about the internal structure of $\mathcal{A}_\psi$, $\mathcal{A}_\phi$, we are free to assume that $\mathcal{A}_\psi = \ket{\psi}\bra{0} + \ket{\phi'}\bra{1} + \sum_{i \ge 2} \ket{\zeta_i} \bra{i}$, $\mathcal{A}_\phi = \ket{\phi}\bra{0} + \ket{\psi'}\bra{1} + \sum_{i \ge 2} \ket{\zeta_i} \bra{i}$. Here $\ket{\phi'}$ and $\ket{\psi'}$ are states orthonormal to $\ket{\psi}$ and $\ket{\phi}$, respectively, within the subspace spanned by $\ket{\psi}$ and $\ket{\phi}$, and $\ket{\zeta_i}$ are arbitrary states which are orthonormal to both of these states and each other. Explicitly, we can take
\[ \ket{\phi'} = \frac{\ket{\phi} - \ip{\psi}{\phi}\ket{\psi}}{\sqrt{1 - |\ip{\psi}{\phi}|^2}},\;\; \ket{\psi'} = \frac{\ket{\psi} - \ip{\phi}{\psi}\ket{\phi}}{\sqrt{1 - |\ip{\psi}{\phi}|^2}}. \]
Then
\[ \| \mathcal{A}_\psi - \mathcal{A}_\phi\| = \|(\ket{\psi}-\ket{\phi})\bra{0} + (\ket{\phi'}-\ket{\psi'})\bra{1}\|. \]
Writing $\ket{\delta} := \ket{\psi}-\ket{\phi}$, $\ket{\delta'} := \ket{\phi'}-\ket{\psi'}$ and upper-bounding the operator norm by the Frobenius norm, we have
\begin{align*}
\| \mathcal{A}_\psi - \mathcal{A}_\phi\| &\le \sqrt{\tr (\mathcal{A}_\psi^\dag - \mathcal{A}_\phi^\dag)(\mathcal{A}_\psi - \mathcal{A}_\phi)}\\
&= \sqrt{\ip{\delta}{\delta} + \ip{\delta'}{\delta'}}\\
&= \sqrt{2} \| \ket{\psi}-\ket{\phi} \|,
\end{align*}
where we use the fact (which can easily be seen by direct calculation) that $\|\ket{\phi'}-\ket{\psi'}\| = \|\ket{\psi}-\ket{\phi}\|$. Hence
\[ \| \mathcal{A}_\psi - \mathcal{A}_\phi\| \le \sqrt{2} \epsilon \]
and in turn, by induction,
\[ \| \ket{\eta}_{\psi,T} - \ket{\eta}_{\phi,T} \| \le T\sqrt{2} \epsilon. \]
As the algorithm is supposed to output something different if it is given  $\mathcal{A}_\phi$ rather than $\mathcal{A}_\psi$, assuming that it succeeds, the final measurement made distinguishes between the two states $\ket{\eta}_{\psi,T}$ and $\ket{\eta}_{\phi,T}$. The optimal worst-case probability $p$ of distinguishing these states is given by the trace distance between them~\cite{helstrom76}, so
\begin{eqnarray*}
p &=& \frac{1}{2} + \frac{1}{4} \| \eta_{\psi,T} - \eta_{\phi,T} \|_1 \le \frac{1}{2} + \frac{1}{2} \| \ket{\eta}_{\psi,T} - \ket{\eta}_{\phi,T} \|\\
&\le& \frac{1}{2} + \frac{T\epsilon}{\sqrt{2}}.
\end{eqnarray*}
Therefore, in order for the algorithm to succeed with probability (say) $2/3$, it must use $\mathcal{A}_\psi$ at least $\Omega(1/\epsilon)$ times. As a simple example of how this bound can be applied, consider an algorithm which attempts to distinguish between these two cases: a) the output from the QLE subroutine is a particular state $\ket{\psi_0}$; b) the output from the QLE subroutine is some state $\ket{\phi}$ such that the overlap $|\ip{\phi}{\psi_0}|^2 = 1-\epsilon$. Then $\| \ket{\phi} - \ket{\psi_0} \| = O(\sqrt{\epsilon})$, so any algorithm distinguishing between these two cases by using QLE as a black box must use it $\Omega(1/\sqrt{\epsilon})$ times.

This bound is tight for this particular problem, which can be solved by using the QLE subroutine $O(1/\sqrt{\epsilon})$ times within quantum amplitude estimation~\cite{brassard02}. However, for other problems it may be possible to put stronger lower bounds on the complexity.





\subsection{Replacing the QLE subroutine with a classical algorithm}

The above lower bound shows, roughly speaking, that any algorithm which uses the QLE subroutine as a black box and attempts to determine up to accuracy $\epsilon$ some property of the output state must make $\Omega(1/\sqrt{\epsilon})$ uses of the subroutine. However, in some cases it can be of interest to approximate properties of the output state to quite low levels of accuracy.

For example, consider the problem of distinguishing between the following two cases: a) the solution to a BVP is periodic; b) the solution is far from periodic. As it is known that quantum algorithms can test periodicity of functions exponentially faster than classical algorithms can~\cite{chakraborty13a}, one might hope to use QLE, together with the quantum periodicity tester, to solve this problem exponentially faster than any classical algorithm.

Also note that it is likely to be hard to prove that it is impossible to obtain a superpolynomial quantum speedup for solving BVPs, if we define ``solving'' a BVP as computing an arbitrary function of the solution to a BVP. For example, we could contrive a BVP where the solution is easy to write down, and can be interpreted as an integer; and could then ask the algorithm to output the prime factors of that integer. Proving that quantum computers could not outperform classical computers for this task would imply an efficient classical algorithm for integer factorisation.

Nevertheless we believe that, even given a quantum algorithm for solving problems of this form, any uses of the QLE algorithm as a subroutine could be replaced with a classical algorithm, with at most a polynomial slowdown if the spatial dimension is fixed and the solution is suitably smooth. This would imply that any exponential quantum speedup in the overall algorithm is not due to the part of it that solves the FEM. Making this argument rigorous seems challenging for technical reasons related to regularity of meshes and comparing different norms to measure accuracy, so we do not attempt it here, instead merely sketching the ideas informally.

The argument proceeds as follows. Imagine we have an overall quantum algorithm which uses the QLE algorithm as a subroutine to solve $T$ FEM instances in spatial dimension bounded by $d = O(1)$, such that the solution to each instance has all relevant Sobolev norms bounded by $O(1)$. Then each such instance can be approximately solved by a classical algorithm using a mesh of size $\poly(1/\epsilon)$, for any desired accuracy $\epsilon$. We replace each subroutine which applies the QLE algorithm to solve an instance of the FEM, using a mesh $\mathcal{M}$ to achieve accuracy $\epsilon$, with the following procedure:
\begin{enumerate}
\item Classically solve the same FEM instance, using a mesh $\mathcal{M}'$ which achieves accuracy $\max\{\gamma/T, \epsilon\}$ for some universal constant $\gamma$. Note that if $\epsilon < \gamma/T$ this will in general be a coarser mesh than $\mathcal{M}$.
\item Construct the quantum state corresponding to the output of the solver, as a superposition of basis functions from $\mathcal{M}'$.
\item Map this quantum state to the equivalent quantum state on the finer mesh $\mathcal{M}$. This is essentially equivalent to the classical task of expressing each element of $\mathcal{M}'$ in terms of elements of $\mathcal{M}$.
\end{enumerate}
Here we are assuming that the meshes $\mathcal{M}$ and $\mathcal{M}'$ are sufficiently regular that the last step makes sense (in particular, that $\mathcal{M}$ is a submesh of $\mathcal{M}'$).

If $\epsilon \ge \gamma/T$, the state produced by the original subroutine is left essentially unchanged. If $\epsilon < \gamma/T$, the original state produced was within distance $O(1/T)$ of the actual solution to the corresponding FEM instance, as is the state produced by the new subroutine. By the triangle inequality, the new state must be within distance $O(1/T)$ of the old state. If each such state produced by one of the new subroutines is within Euclidean distance $O(1/T)$ of the corresponding original state produced by one of the QLE subroutines, then using a similar argument to Section \ref{sec:lower1}, the whole algorithm does not notice the difference between the original and modified sequence of subroutines except with low probability.

We now examine the complexity of the steps in the modified subroutines. Each use of step 1 solves the FEM with precision $O(1/T)$, which requires time $\poly(T)$ and a mesh of size $\poly(T)$. In step 2 we need to construct a known $\poly(T)$-dimensional quantum state. This can be done in time $\poly(T)$ for any such state (see e.g.~\cite[Claim 2.1.1]{prakash14}). If $\mathcal{M}$ and $\mathcal{M}'$ are suitably regular, the mapping  required for step 3 can be implemented efficiently, i.e.\ in time polynomial in $n$, the number of qubits used by the original algorithm.

As the original quantum algorithm solved $T$ instances of the FEM, and acts nontrivially on all $n$ qubits, its runtime must be lower-bounded by $\max\{T,n\}$. Therefore, the runtime of the new algorithm is at most polynomial in the runtime of the old algorithm. As the new algorithm no longer contains any quantum subroutines which solve the FEM, we see that any quantum speedup achieved by it does not come from quantum acceleration of the FEM.


\subsection{Solving oracular FEM instances}
\label{sec:oracular}

We finally observe that there cannot be an efficient quantum (or classical) algorithm for solving an instance of the FEM if the input function $f(x)$ is initially unknown and provided via an oracle (``black box''), and does not satisfy some smoothness properties. Indeed, this even holds for near-trivial FEM instances.

Imagine we are given an FEM instance of the form $u(x) = f(x)$, for $f \in L^2[0,1]$, and are asked to approximate the quantity $\int_0^{\frac{1}{2}} u(x)^2 dx$ to within accuracy $\epsilon$ -- this is a very simple property of a trivial PDE. Further assume that we are given access to $f$ via an oracle which maps $x \mapsto f(x)$ for $x \in [0,1]$, and that there are $N$ possibilities for what the function $f$ can be. We will show that this problem is hard by encoding unstructured search on $N$ elements as an instance of the FEM.

Let $B$ be the ``bump'' function defined by $B(x) = \exp(-1/(1-x^2))$ for $-1 < x < 1$, and $B(x) = 0$ elsewhere. Fix $N$ and let $f_0$ be the shifted and rescaled bump function $f_0(x) = \sqrt{N} B(2Nx-1)$. $f_0$ is supported only on $[0,1/N]$, has continuous derivatives of all orders, and $\| f_0 \| = \Theta(1)$.


Assume we have access to an oracle function $O:\{0,\dots,N-1\} \rightarrow \{0,1\}$ such that there is a unique $y_0 \in \{0,\dots,N-1\}$ with $O(y_0) = 1$. It is known that determining whether $y_0 < N/2$ or $y_0\ge N/2$ requires $\Omega(\sqrt{N})$ quantum queries to $O$~\cite{grover05}. We define $f$ in terms of $O$ as follows. Given $x \in [0,1]$, set $y = \lfloor N x \rfloor$ and evaluate $O(y)$. If the answer is 1, return $f_0(x-y/N)$. Otherwise, return 0.

$f$ (equivalently, $u$) is a bump function on the range $[y_0/N,(y_0+1)/N]$, and is zero elsewhere. So, if $y_0 < N/2$, $\int_0^{\frac{1}{2}} u(x)^2 dx \ge C$ for some constant $C>0$, while if $y_0 \ge N/2$, $\int_0^{\frac{1}{2}} u(x)^2 dx = 0$. Hence approximating this integral up to additive accuracy $\epsilon$, for sufficiently small constant $\epsilon > 0$, allows us to determine whether or not $y_0 < N/2$. As this task requires $\Omega(\sqrt{N})$ quantum queries, solving this instance of the FEM must require $\Omega(\sqrt{N})$ queries to $f$. A similar classical lower bound of $\Omega(N)$ queries also holds. Note that this does not contradict the bound (\ref{eq:disc_error}) as the norms of derivatives of $u$ are large.


\section{Conclusions}
\label{sec:conclusions}

We have shown that, when one compares quantum and classical algorithms for the FEM fairly by considering every aspect of the problem -- including the complexity of producing an accurate approximation of the desired classical output -- an apparent exponential quantum advantage can sometimes disappear. However, there are still two types of problem where quantum algorithms for the FEM could achieve a significant advantage over classical algorithms: those where the solution has large higher-order derivatives, and those where the spatial dimension is large.

For ease of comparison with the quantum algorithm, we have only considered a very simple classical FEM algorithm here; there is a large body of work concerned with improving the complexity of such algorithms. For example, the finite element mesh can be developed adaptively and made more refined near parts of the domain which are more complex or of particular interest. This can substantially improve the convergence speed. It is our suspicion that more advanced classical FEM algorithms might eliminate the quantum algorithm's advantage with respect to BVPs whose solutions have large higher-order derivatives.

For example, adaptive schemes such as ``hp-FEM'' have, in principle, a discretisation error that scales far better than the scaling shown here; it can be shown~\cite{guo86} that a perfect adaptive scheme has scaling
\[ \Vert u - \tilde{u} \Vert = O(e^{-{\sfrac{1}{h}}}),\]
provided that the dimension of the domain is both small and fixed. While this is a large improvement over the ``vanilla'' classical complexity presented above, it is not always apparent how to generate adaptive schemes that are effective enough to saturate this scaling, in practice. Also, it does not seem impossible that the quantum algorithm could be substantially improved using similar adaptive schemes.

Additionally, the case for the possibility of substantial improvement in the classical algorithm is less clear with respect to problems in high spatial dimension $d$. Indeed, any reasonable discretisation procedure seems likely to lead to systems of linear equations which are of size exponential in $d$ (this is the so-called ``curse of dimensionality''). This is precisely the regime in which the quantum algorithm might be expected to have a significant advantage. One setting in which such high-dimensional BVPs occur is mathematical finance; for example, the problem of pricing multi-asset basket options using the Black-Scholes equation~\cite{jiang05}. Alternatively, producing a solution to any problem in many-body dynamics requires solving a PDE where the dimension grows with the number of bodies. However, Monte Carlo methods and related techniques can sometimes be used to alleviate the curse of dimensionality in practice~\cite{boyle97,lecuyer09}. It is therefore an interesting open question whether quantum algorithms can in fact yield an exponential speedup for problems of practical interest in this area.


\subsection*{Acknowledgements}

AM is supported by the UK EPSRC under Early Career Fellowship EP/L021005/1 and would like to thank Robin Kothari for helpful discussions and explanations of the results in~\cite{childs15b}. SP is supported by the EPSRC Centre for Doctoral Training in Quantum Engineering.

\appendix


\section{Use of HHL for approximating the norm of the solution}
\label{sec:hhlnorm}

Assume that we have an $s$-sparse system of linear equations $A \mathbf{x} = \mathbf{b}$, for some Hermitian $N \times N$ matrix $A$ such that $\lambda_{\max}(A) \le 1$, $\lambda_{\min}(A) \ge 1/\kappa$. We would like to approximate $\|\mathbf{x}\|$ up to accuracy $\epsilon \|\mathbf{x}\|$ using the HHL algorithm~\cite{harrow09}. Here we sketch how the complexity of this task can be bounded, using the same notation as Theorem \ref{thm:cks} (see~\cite{harrow09} for further technical details). The HHL algorithm is based on a subroutine $\mathcal{P}_{\text{sim}}$ whose probability of acceptance is approximately $p := \| A^{-1} \ket{b} \|^2 / \kappa^2$. For any $\delta>0$, approximating the probability $p$ that a subroutine accepts, up to additive accuracy $\delta p$, can be achieved using amplitude estimation~\cite{brassard02} with $O(1/(\delta \sqrt{p}))$ uses of the subroutine. Therefore, approximating $\kappa \|\mathbf{b}\| \sqrt{p} = \|\mathbf{x}\|$ up to additive accuracy $\epsilon \|\mathbf{x}\|$ can be achieved with
\[ O\left(\frac{\kappa \|\mathbf{b}\|}{\epsilon\|\mathbf{x}\|}\right) = O\left(\frac{\kappa}{\epsilon}\right) \]
uses of $\mathcal{P}_{\text{sim}}$, where we use $\lambda_{\max}(A) \le 1$. The runtime of the $\mathcal{P}_{\text{sim}}$ subroutine, which is described in~\cite{harrow09}, depends on the accuracy with which its actual probability of acceptance $\widetilde{p}$ approximates $p$. Using the best known algorithm for Hamiltonian simulation~\cite{berry15} within $\mathcal{P}_{\text{sim}}$, an accuracy of $|\widetilde{p}-p| = O(\epsilon p)$ can be achieved with $O((s \kappa/\epsilon) \poly \log(s\kappa/\epsilon))$ uses of the algorithm $\mathcal{P}_A$ for determining entries of $A$. The runtime is the same up to a polylogarithmic term in $N$, $s$, $\kappa$, and $\epsilon$. Each use of the subroutine within amplitude estimation requires two uses of $\mathcal{P}_b$ to reflect about the state $\ket{b}$. Therefore, the overall number of uses of $\mathcal{P}_A$ required is
\[ O((s \kappa^2/\epsilon) \poly \log(s\kappa/\epsilon)), \]
and the number of uses of $\mathcal{P}_b$ is $O(\kappa/\epsilon)$. Note that quantum linear equations algorithms subsequent to HHL~\cite{ambainis12b,childs15b} achieved better dependence on $\kappa$, $\epsilon$, or both for the task of producing $\ket{x}$; however, it does not seem obvious how to use these to achieve improved accuracy for estimating $\|\mathbf{x}\|$.


\section{Proof of technical bound}
\label{sec:techbounds}

In this appendix we prove the claimed bound in Section \ref{sec:overall} that
\[ \frac{\alpha}{\|r\|} = O(h \sqrt{s}), \]
where $\alpha = \left(\sum_i \langle \phi_i,r\rangle^2 \right)^{1/2}$. Indeed, we show that
\[ \sup_{r \neq 0} \frac{\left(\sum_i \langle \phi_i,r\rangle^2 \right)^{1/2}}{\|r\|} = O(h \sqrt{s}).
\]
Observe that this expression will be maximised when $r$ is in the subspace spanned by the $\{\phi_i\}$ functions, so we can assume that $r = \sum_i \mathbf{r}_i \phi_i$ for some $\mathbf{r}_i$. Then the numerator satisfies
\begin{eqnarray*}
\left(\sum_i \langle \phi_i,r\rangle^2 \right)^{1/2} &=& \left( \sum_i \left(\int_\Omega \phi_i(\mathbf{x}) r(\mathbf{x}) d\mathbf{x} \right)^2 \right)^{1/2}\\
&=& \left( \sum_i \left(\int_\Omega \phi_i(\mathbf{x}) \sum_j \mathbf{r}_j \phi_j(\mathbf{x}) \right)^2 \right)^{1/2}\\
&=& \left( \sum_i \left( \sum_j \mathbf{r}_j \int_\Omega \phi_i(\mathbf{x}) \phi_j(\mathbf{x}) d\mathbf{x}\right)^2 \right)^{1/2}\\
&=& \| W \mathbf{r}\|,
\end{eqnarray*}
where we define the matrix $W_{ij} := \int_\Omega \phi_i(\mathbf{x}) \phi_j(\mathbf{x}) d\mathbf{x}$. Similarly, for the denominator we have
\begin{eqnarray*} \|r\| &=& \left(\int_\Omega \left(\sum_i \mathbf{r}_i \phi_i(\mathbf{x}) \right)^2 d\mathbf{x} \right)^{1/2}\\
&=& \left(\sum_{i,j} \mathbf{r}_i \mathbf{r}_j \int_\Omega \phi_i(\mathbf{x}) \phi_j(\mathbf{x}) d\mathbf{x} \right)^{1/2}\\
&=& (\mathbf{r}^T W \mathbf{r})^{1/2}.
\end{eqnarray*}
Therefore,
\begin{eqnarray*} \frac{\alpha}{\|r\|} &\le& \sup_{\mathbf{r} \neq 0} \left( \frac{\mathbf{r}^T W^T W \mathbf{r}}{\mathbf{r}^T W \mathbf{r}} \right)^{1/2}\\
&=& \sup_{\mathbf{r}', \|\mathbf{r}'\| = 1} ((\mathbf{r'})^T W \mathbf{r}')^{1/2}\\
&=& \|W\|^{1/2}.
\end{eqnarray*}
Assume that $W$ is $s$-sparse. To upper-bound $\|W\|$ we use
\[ \|W\| \le s \max_{i,j} |W_{ij}| = s \max_{i,j} |\langle \phi_i, \phi_j \rangle| \le s \max_i \|\phi_i\|^2, \]
where the first inequality can be found in~\cite{childs09b}, for example, and the second is Cauchy-Schwarz. Then
\[ \| \phi_i \|^2 = \int_T \phi_i(\mathbf{x})^2 d\mathbf{x} \le h^d \max_{\mathbf{x} \in T} \phi_i(\mathbf{x})^2 = O(h^2), \]
where we assume that $\phi_i$ is supported on a region $T$ of diameter at most $h$, and we use (\ref{eq:basisnorm}) to bound $\max_{\mathbf{x} \in T} \phi_i(\mathbf{x})^2 = O(h^{2-d})$. Thus $\alpha / \|r\| = O(h \sqrt{s})$.

\bibliographystyle{plain}
\bibliography{thesis,additional}

\end{document}